\pdfoutput=1
\documentclass[prd,10pt,aps,superscriptaddress,floatfix,showkeys,showpacs,twocolumn]{revtex4-1}

\usepackage{xcolor,epsfig,amsmath,amssymb,subfigure,placeins}
\usepackage{graphicx,amsmath,bbm,bm,array}
\usepackage[colorlinks=true,linktocpage=true,linkcolor=blue,citecolor=blue]{hyperref}

\newcommand{\be}{\begin{equation}}
\newcommand{\ee}{\end{equation}}
\newcommand{\ba}{\begin{eqnarray}}
\newcommand{\ea}{\end{eqnarray}}
\newcommand{\nn}{\nonumber\\}

\begin{document}
\title{Collective modes in the anisotropic collisional hot QCD medium at finite chemical potential}
\author{Mohammad Yousuf Jamal}
\email{mohammad@iitgoa.ac.in}
\affiliation{School of Physical Sciences, Indian Institute of Technology Goa, Ponda-403401, Goa, India}

\begin{abstract}
We conducted a study on the collective modes within the hot QCD medium generated in heavy-ion collision experiments. These modes, whether real or imaginary, stable or unstable, play a crucial role in shaping the medium's evolution. To gain a deeper understanding, we considered several factors affecting the medium, including anisotropy, interactions among medium particles, and finite baryonic chemical potential. While the first two aspects have been thoroughly examined from various angles, the inclusion of finite chemical potential was previously overlooked.
To provide a comprehensive analysis, we integrated these factors. The interactions among medium particles were accounted for using the BGK collisional kernel, while anisotropy and finite chemical potential were incorporated through the distribution functions of quarks, anti-quarks, and gluons. Our findings suggest that the presence of finite chemical potential amplifies the influence of unstable modes, potentially affecting the rapid thermalization of the hot QCD medium.
Furthermore, exploring the implications of finite chemical potential in conjunction with other aspects of the created medium is intriguing, particularly in the context of low-energy heavy-ion collision experiments conducted at low temperatures and finite baryon density.
\end{abstract}
\maketitle

\section{Introduction}

In relativistic heavy-ion collision (HIC) experiments, an extremely high-energy density is created, somewhat similar to that existing shortly after the Big Bang. This leads to the formation of quark-gluon plasma (QGP), a phase where quarks, anti-quarks, and gluons are the dominant components, distinct from hadronic matter~\cite{expt_rhic,expt_lhc}. The historic discovery of the $J/\Psi$, a bound state of the charm quark and anti-charm quark, proved the existence of the heavy quarks, emphasized that the nucleus–nucleus collisions at high energy are very different from the simple superposition of nucleon-nucleon interactions. In HIC, the process begins with a high-temperature region resulting from the collision, leading to the formation of the QGP, which then cools and condenses into mesons, baryons, or hadrons as it expands. Finally, the system reaches "freeze-out," where hadrons no longer interact and stream into the detectors.

Various theoretical approaches, including semi-classical transport, kinetic, hard-thermal loop, ASD-CFT, and holographic theories, have been used to study the matter produced in HIC~\cite{Ryu, Denicol1}. Experimental observations, such as the suppression of heavy quarkonia yields due to color screening ~\cite{Chu:1988wh, Agotiya:2016bqr, Jamal:2018mog, Jamal:2020rvh}, Landau damping~\cite{Landau:1984, Jamal:2020hpy} and energy loss or gain of heavy quarks ~\cite{Carrington:2015xca, Ghosh:2023ghi, Koike:1991mf, Jamal:2020fxo, Jamal:2021btg, Jamal:2020emj, YousufJamal:2019pen}, support the formation of QGP. The observables detected at the detector end could be affected by the collective excitations ~\cite{Mrowczynski:1993qm, Mrowczynski:1994xv, Mrowczynski:1996vh, Jamal:2017dqs, Kumar:2017bja,avdhesh} the QGP medium, which can be stable or unstable, playing a significant role in shaping these outcomes.

In many-body systems, the collective excitations spectrum provides crucial information about thermodynamic properties, transport, and non-equilibrium evolution~\cite{Bellac:1996}. Analogous to quantum electrodynamics (QED), the evolving medium in HICs may exhibit instabilities akin to those in electromagnetic plasma (EMP), potentially aiding fast thermalization~\cite{Weibel:1959zz, Mrowczynski:2004kv}. At very high temperatures, QCD resembles QED due to a small coupling constant, and both EMP and QGP display collective behavior, suggesting QGP has a rich spectrum of collective modes similar to EMP~\cite{dm_rev1,dm_rev2}. In the present context, these modes could refer to plasma excitations that could be real or imaginary and stable or unstable. The stable modes have infinite lifetimes, whereas the unstable modes decay quickly. Stable modes exhibit a long-range interaction with heavy quarks moving through the QGP medium, while unstable modes are exclusive to anisotropic plasma and potentially contribute to rapid thermalization and equilibration. This lesser-explored puzzle of unstable modes garners interest, leading to their in-depth investigation, whereas the former stable modes have received less attention in the literature despite extensive discussion of the latter.

In various analyses, the formation and effects of collective modes in plasma have been explored, considering factors like anisotropy, particle collisions, etc~\cite{Romatschke:2003ms, Romatschke:2004jh, Romatschke:2003ms, Romatschke:2004jh, Arnold:2003rq, Mrowczynski:2005ki, Schenke:2006yp,  Schenke:2006xu, Carrington:2014bla, Karmakar:2022one}. We focus on the critical aspect of including finite baryon density in these studies as the QCD phase diagram signals the importance of incorporating it, especially at a low temperature. A high baryon density implies an excess of quarks over antiquarks, characterized by the baryon chemical potential ($\mu$). At zero chemical potential, quark and antiquark densities are equal, suitable for high-energy HIC at Relativistic Heavy Ion Collider (RHIC) and Large Hadron Collider (LHC) and the early Universe, but less so for lower energies. This is vital for experimental facilities operating at moderate temperatures with finite baryon density, such as SPS at CERN~\cite{Doble:2017syb}, FAIR in Darmstadt, Germany\cite{Selyuzhenkov:2020djo}; NICA in Dubna, Russia~\cite{Syresin:2022mjz}; and J-PARC in Tōkai, Japan~\cite{Nagamiya:2006en}. Our comprehensive study of collective modes considers medium particle collisions through BGK - collisional kernel, whereas momentum anisotropy and finite chemical potential through particle distribution functions in the Boltzmann transport equation. Dispersion relations for these modes are obtained from the poles of the gluon propagator. Their solutions allow us to identify the characteristics of these modes in detail.

The manuscript is organized as follows: In Section~\ref{GSE}, we discuss the methodology for obtaining mode dispersion relations. Section~\ref{RD} contains a detailed results discussion, and Section~\ref{SC} summarizes our findings, conclusions, and potential future research. Throughout the text, natural units are used with $c=k_B=\hbar=1$. We denote three vectors in bold and four vectors in regular font. The center dot represents the four-vector scalar product with $g_{\mu\nu}={\text {diag}}(1,-1,-1,-1)$.

\section{Methodology}
\label{GSE}

Collective modes in the QGP medium arise when an initially homogeneous and stationary state is perturbed by fluctuations or an external field, leading to local charges, currents, and chromo-electric and chromo-magnetic fields. These fields interact with colored partons, giving rise to collective motion if the perturbation wavelength exceeds the inter-particle spacing. The nature of these modes is determined by solutions to their dispersion relations ($\omega({\bf k})$, with {\bf k} as the wave vector), obtained from the poles of the gluon propagator. Complex-valued solutions indicate stability or instability: $\Im(\omega({\bf k}))=0$ represents stable modes, $\Im(\omega({\bf k}))>0$ signifies instability with exponential growth ($e^{\Im(\omega({\bf k}))t}$), and $\Im(\omega({\bf k}))<0$ implies damped modes with exponential decay ($e^{-\Im(\omega({\bf k}))t}$). Modes with $-\Im(\omega({\bf k}))\geq |\Re(\omega({\bf k}))|$ are overdamped. Next, we obtain the gluon propagator that involves deriving the gluon polarization tensor, which encapsulates medium-specific information like anisotropy, finite chemical potential, and medium particle collision effects.

\subsection{Gluon polarization tensor}

The gluon polarization tensor ($\Pi^{\mu\nu}$) serves as a repository of information regarding the QCD medium and acts as the interaction term in the effective action of QCD. {We have chosen to focus on the soft momentum scale, where collective motion in the hot medium first becomes apparent ($k \sim gT \ll T$). At this scale, field fluctuations are represented by $A^{\mu} \sim O(\sqrt{g}T)$, and derivatives are of the order $\partial_x \sim O(gT)$ \cite{Blaizot:2001nr}. The expression for the non-abelian field strength tensor is represented as:
\begin{equation}
F^{\mu\nu}=\partial^{\mu }A^{\nu}-\partial^{\nu }A^{\mu }-i g\left[A^{\mu },A^{\nu }\right].
\label{eq:fmunu}
\end{equation}
Considering the soft momentum scale, the $F^{\mu\nu}$ modifies, and our attention pivots towards the abelian part, as the non-abelian component in Eq.\eqref{eq:fmunu} comes at an order of O($g^2$), diminishes significantly in comparison to the first two terms, which are at an order of O($g^{3/2}$). 
Now, let $\delta f^{i}_a(p,X)$ be the small perturbation in the equilibrium distribution functions $f^{i}(\mathbf{p})$. The linearized semi-classical transport equations can be written separately for each color
channel as~\cite{Elze:1989un, Mrowczynski:1993qm, Romatschke:2003ms, Jiang:2016dkf,Schenke:2006xu},
\ba
v^{\mu}\partial_{\mu}\delta f^{i}_a(p,X) + g \theta_{i}v_{\mu}F^{\mu\nu}_a(X)\partial_{\nu}^{(p)}f^{i}(\mathbf{p}) = \mathcal{C}^{i}_a(p,X),
\label{eq:Bol}
\ea
such that
\ba f^{i}_a(p,X)=f^{i}(\mathbf{p})+\delta f^{i}_a(p,X), 
\ea
with
\ba\delta f^{i}_a(p,X)\ll f^{i}(\mathbf{p}),
\ea
where $a$ is the color index, and $i$ refers to the particle species (quark, anti-quark, or gluon). $\theta_{i}$ takes values as $\theta_{g}=\theta_{q}= 1$  and $\theta_{\bar{q}}= - 1$. The four vectors $x^{\mu}=(t,\mathbf{x}) = X$ and $v^{\mu}=(1,\mathbf{v}=\mathbf{p}/|\mathbf{p}|) = V$ represent space-time coordinates and the velocity of the plasma particle, respectively. On solving Eq.\eqref{eq:Bol}, $\delta f^{i}_a(p,X)$ can be obtained that gives rise to the induced current in the medium,
\ba
J_{\text{ind}, a}^{\mu}(X) &=& g\int\frac{d^{3}p}{(2\pi)^3} v^{\mu}\{2N_c \delta f^{g}_a(p,X)+N_{f}[\delta f^{q}_a(p,X)\nn 
&-&\delta f^{\bar{q}}_a(p,X)]\}
\label{eq:jin}
\ea
Now, in the Fourier space, the gluon polarization tensor, $\Pi^{\mu\nu}_{ab}(K)$ can be calculated from the induced current, $J_{\text{ind}, a}^{\mu}(K)$ using the relation \cite{Mrowczynski:2000ed},
\ba
\Pi^{\mu\nu}_{ab}(K) = \frac{\delta J_{\text{ind}, a}^{\mu}(K)}{\delta A_{\nu}^b(K)},
\label{eq:pimunu}
\ea
where, $k^{\mu} = (\omega,{\bf k})\equiv K$. Next, to consider the presence of medium particle collisions, the collisional term $\mathcal{C}^{i}_a(p, X)$ given in Eq.\eqref{eq:Bol} is described by the BGK-kernel ~\cite{Bhatnagar:1954, Jiang:2016dkf, Carrington:2004} that depicts the equilibration of the system due to the collisions in a  time proportional to $\nu^{-1}$ and is given as,
\ba
\mathcal{C}^{i}_a(p,X)=-\nu\left[f^{i}_a(p,X)-\frac{N^{i}_a(X)}{N^{i}_{\text{eq}}}f^{i}_{\text{eq}}(|\mathbf{p}|)
\right],
\ea
where $N^{i}_a(X)$ is the particle number, and $N^{i}_{\text{eq}}$ is its equilibrium value given as,
\ba
\label{particlenumber1}
N^{i}_a(X)=\int \frac{d^{3}p}{(2\pi)^3} f^{i}_a(p,X)\text{ , ~} \\ N^{i}_{\text{eq}} = \int \frac{d^{3}p}{(2\pi)^3} f^{i}_{\text{eq}}(|\mathbf{p}|).\label{particlenumber2}
\ea

 At finite baryon density or quark chemical potential ($\mu$), the medium particle distribution functions  are given as
\ba
f_{g}({\bf p}) = \frac{\exp[-\beta E_g]}{1-\exp[-\beta E_g]},
\ea
\ba
f_{q/{\bar q}}({\bf p}) = \frac{\exp[-\beta ( E_q\mp \mu)]}{1+ \exp[-\beta ( E_q\mp \mu)]}.
\ea
where, $E_{g}=|{\bf p}|$ for the gluons and, $ E_{q/{\bar q}}=\sqrt{|{\bf p}|^2+m_q^2}$ for the quark degrees of freedom ($m_q$, denotes the mass of the light quarks).
Since, $m_q \ll T$, the mass of the light quarks can be neglected as compared to 
the temperature of the medium. The effective distribution function is defined as, 
\begin{equation}
f(p)=2N_c f^g(\mathbf{p})+N_f\left[f^q(\mathbf{p})+f^{\bar{q}}(\mathbf{p})\right]\text{,}
\end{equation}
and that in equilibrium is given as,
\begin{equation}
f_{\text{eq}}(|\mathbf{p}|)=2N_c f_{\text{eq}}^{g}(|\mathbf{p}|)+N_f\left[f_{\text{eq}}^q(|\mathbf{p}|)+f_{\text{eq}}^{\bar{q}}(|\mathbf{p}|)\right],
\end{equation}
where $N_c$ is the number of color charges, and $N_f$ is the number of flavors.
Next, in the presence of momentum anisotropy, the isotropic distribution function $f^{i}(\mathbf{p})$ is rescaled (stretched and squeezed) in one direction in the momentum space as follows \cite{Romatschke:2003ms, Romatschke:2004jh},
\ba
f({\mathbf{p}}) \equiv f_{\xi}({\mathbf{p}}) = f(\sqrt{{\bf p}^{2} + \xi({\bf p}\cdot{\bf \hat{n}})^{2}}),
\label{eq:aniso}
\ea
where ${\mathbf{\hat{n}}}$ is a unit vector showing the direction of momentum anisotropy, and $\xi$ is the strength of anisotropy. Now using $f({\mathbf{p}})$ from Eq.\eqref{eq:aniso}, we solve Eq.\eqref{eq:Bol} for $\delta f^{i}_a(p,K)$ in the Fourier space we obtained \cite{Kumar:2017bja}, }
\begin{widetext}
\ba
\delta f^{i}_a(p,K)=\frac{-ig\theta_iv_{\mu}F_a^{\mu\nu}(K)\partial_{\nu}^{(p)}f^{i}(\mathbf{p})+i\nu(f^{i}_{\text{eq}}(|\mathbf{p}|)-f^{i}(\mathbf{p}))+i\nu f^{i}_{\text{eq}}(|\mathbf{p}|)\left( \int\frac{d^{3}p^{\prime}}{(2\pi)^3}\delta f_a^{i}(p^{\prime},K) \right)/N_{\text{eq}}}{\omega-\mathbf{v}\cdot\mathbf{k}+i\nu}\text{,} \label{induceddistribution}
\ea
\end{widetext}
where $\delta f^{i}(p,K)$ and $F^{\mu\nu}(K)$, are the
Fourier transforms of $\delta f^{i}(p,X)$ and
$F^{\mu\nu}(X)$, respectively. Using Eq.\eqref{induceddistribution} in Eq.\eqref{eq:jin} (in Fourier space) and solving Eq.\eqref{eq:pimunu} for $\Pi^{\mu\nu}_{ab}(K)$, we obtain,
\begin{widetext}
\ba
\Pi^{\mu\nu}_{ab}(K)&=&\delta_{ab} g^2 \int\frac{d^{3}p}{(2\pi)^3} v^{\mu}\partial_{\beta}^{(p)}f(\mathbf{p})\mathcal{M}^{\beta\nu}(K,V)D^{-1}(K,\mathbf{v},\nu)+\delta_{ab} g^2 (i \nu)\int \frac{d\Omega}{4\pi}v^{\mu}\nn
&\times& D^{-1}(K,\mathbf{v},\nu)\int\frac{d^{3}p^{\prime}}{(2\pi)^3}\partial_{\beta}^{(p^{\prime})}f(\mathbf{p}^{\prime})\mathcal{M}^{\beta\nu}(K,V^{\prime})D^{-1}(K,\mathbf{v}^{\prime},\nu)\mathcal{W}^{-1}(K,\nu),
\label{selfenergy}
\ea
\end{widetext}
where, 
\ba
D(K,\mathbf{v},\nu)=\omega+i\nu-\mathbf{k}\cdot\mathbf{v},
\ea
\ba
\mathcal{M}^{\nu\alpha}(K,V)=g^{\nu\alpha}(\omega-\mathbf{k}\cdot\mathbf{v})-K^{\nu}v^{\alpha}, 
\ea
\ba
\mathcal{W}(K,\nu)=1-i \nu \int \frac{d\Omega}{4\pi}D^{-1}(K,\mathbf{v},\nu),
\ea
Rewriting Eq.~\eqref{selfenergy}, in temporal gauge for anisotropic hot QCD medium,
\begin{widetext}
\ba
\Pi^{ij}(K)&=&m_D^2(\mu, T)~ \int\frac{d\Omega}{4\pi}v^i\frac{v^l+\xi(\mathbf{v}\cdot\mathbf{\hat{n}})n^l}{(1+\xi(\mathbf{v}\cdot\mathbf{\hat{n}})^2)^2} \left[\delta^{jl}(\omega-\mathbf{k}\cdot\mathbf{v})+v^jk^l\right]D^{-1}(K,\mathbf{v},\nu)+(i\nu) m_D^2(\mu, T)~ \int\frac{d\Omega^{\prime}}{4\pi}(v^{\prime})^i\nn
&\times& D^{-1}(K,\mathbf{v}^{\prime},\nu) \int\frac{d\Omega}{4\pi}\frac{v^l+\xi(\mathbf{v}\cdot\mathbf{\hat{n}})n^l}{(1+\xi(\mathbf{v}\cdot\mathbf{\hat{n}})^2)^2}\big[\delta^{jl}(\omega-\mathbf{k}\cdot\mathbf{v})+v^jk^l\big] D^{-1}(K,\mathbf{v},\nu)\mathcal{W}^{-1}(K,\nu),
\label{pimunu}
\ea
\end{widetext}
where the squared Debye mass is given as,
{\ba
m^{2}_D &=& 4\pi \alpha_s(\mu, T) \bigg(-2N_c \int \frac{d^3 p}{(2\pi)^3} \partial_p f_g({|\bf p|})\nn &-& N_f \int \frac{d^3 p}{(2\pi)^3} \partial_p \left(f_q({|\bf p|})+f_{\bar q}({|\bf p|})\right)\bigg),
\label{eq:m} 
\ea  }
where, $m^{2}_D\equiv m^{2}_D(\mu, T)$. Now, solving Eq.~\eqref{eq:m}, we obtained
\ba
m^{2}_D &=& \frac{2 \alpha_s (\mu ,T)}{3 \pi } \left(\pi ^2 T^2 (2 N_c+N_f)+3 \mu ^2 N_f\right)\nn
\label{eq:mdh}
\ea
and the strong coupling constant, $\alpha_s(\mu ,T)$ at finite temperature and chemical potential is given as ~\cite{Srivastava:2010xa, Braaten:1991gm, Bannur:2007tk},
{\small \ba
\alpha_{s}(\mu, T)&=&\frac{g^2_{s}(\mu, T)}{4 \pi}\nn
&=& \frac{6\pi}{\left(33-2N_f\right)\ln\left(\frac{T}{\Lambda_T} \sqrt{1+\frac{\mu^2}{\pi^2 T^2}}\right)}\nn
&\times& \bigg(1-\frac{3\left(153-19N_f\right)}{\left(33-2N_f\right)^2}
\frac{\ln \left(2\ln \left( \frac{T}{\Lambda_T} \sqrt{1+\frac{\mu^2}{\pi^2 T^2}}\right)\right)}{\ln \left( \frac{T}{\Lambda_T} \sqrt{1+\frac{\mu^2}{\pi^2 T^2}}\right)}\bigg).\nn
\label{eq:alpha_s}
\ea}
where $\Lambda_T\ll T$ is the QCD scale parameter. 

\begin{figure}[ht!]  %
	\includegraphics[height=6cm,width=7cm]{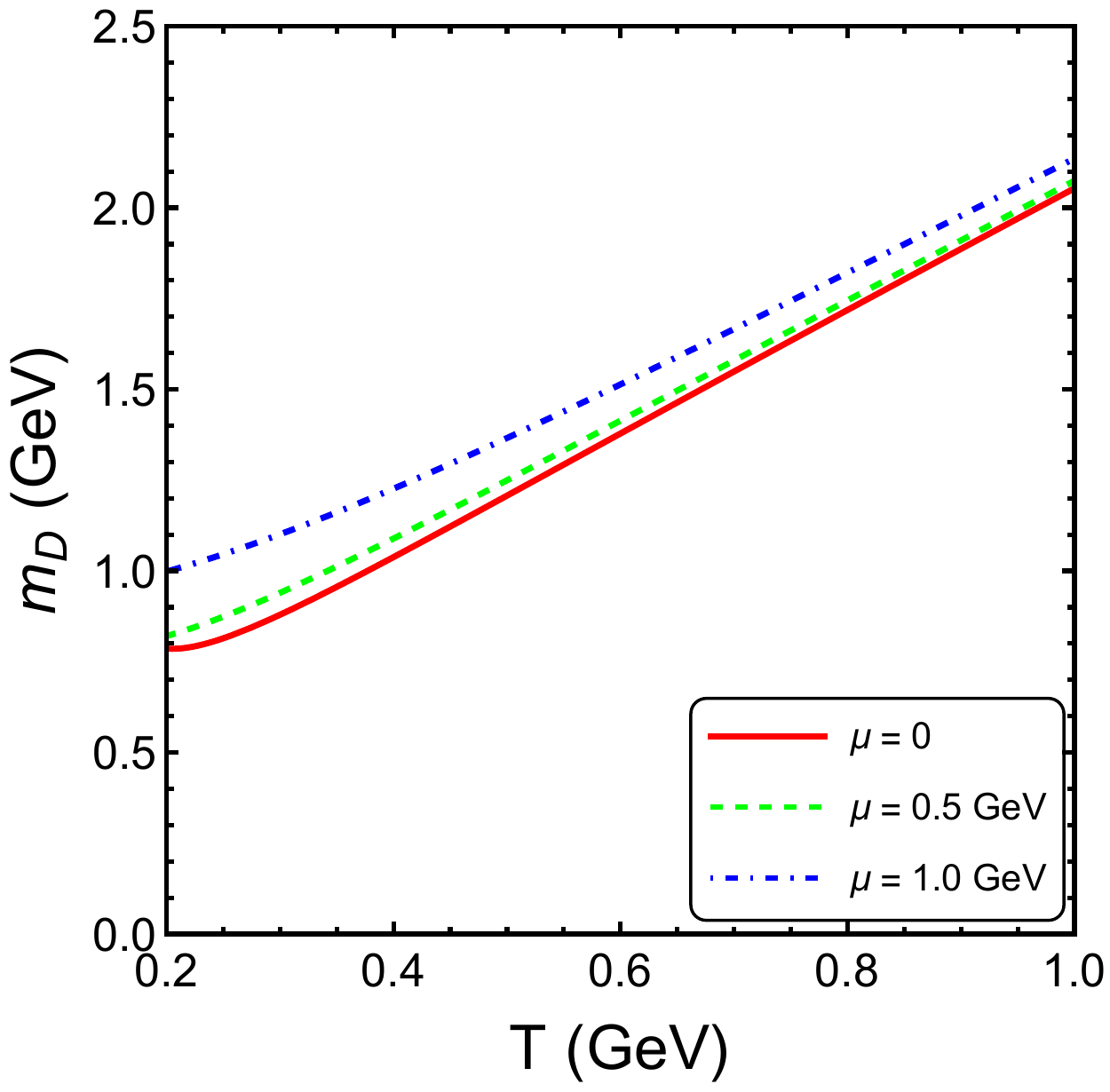}
	\caption{{Variation of screening mass with respect to temperature at fixed values of chemical potential, $\mu=0,0.5,1.0$ GeV}.}
	\label{fig:mD}
\end{figure}

\subsection{Gluon Propagator}
To obtain the gluon propagator, we initiate with Maxwell's equation in the Fourier space,
\ba
i k_{\mu}F^{\mu\nu}(K)=J^{\nu}_{ind}(K)+J^{\nu}_{ext}(K), 
\label{Maxwell:1}
\ea
where, $J^{\nu}_{ext}(K)$ is the external current. { Using the dispersion relation~\cite{Schenke:2006xu},
\ba
J^{\nu}_{ind}(K)= \Pi^{\mu\nu}(K)A_{\mu}(K),
\ea
and $F^{\mu\nu}(K)$ in the Fourier space using Eq.\eqref{eq:fmunu}, Eq.\eqref{Maxwell:1} can be read as,
\ba
[K^{2}g^{\mu\nu}-k^{\mu}k^{\nu}+\Pi^{\mu\nu}(K)]A_{\mu}(K)=J^{\nu}_{ext}(K). 
\ea
The quantity in the square bracket represents the inverse of the gluon propagator. Now, assuming the temporal gauge, $A_0=0$ ( with $ A^{j} = \frac{E^{j}}{i\omega}$), the above equation can be obtained as,
\ba
[\Delta^{-1}(K)]^{ij}E^{j}&=&
i\omega{J^{i}_{ext}}(k), 
\label{eqinA0gauge}
\ea}
where $E^{j}$ is the physical electric field and
\ba
[\Delta^{-1}(K)]^{ij}=({k^2}-{\omega}^{2})\delta^{ij}-k^{i}k^{j}+\Pi^{ij}(K),
\label{invprop}
\ea
is the inverse of the propagator. The dispersion equations for collective modes can be obtained from the 
 poles of the propagator, $[\Delta(K)]^{ij}$. To do so, we first need to obtain, $[\Delta(K)]^{ij}$ from its inverse. For that we begin with $\Pi^{ij}$ by constructing an analytical form of it using the available symmetric tensors such as,
\ba
\Pi^{ij}=\alpha{P}^{ij}_{T}+\beta{P}^{ij}_{L}+\gamma{P}^{ij}_{n}+\delta{P}^{ij}_{kn},
\label{seexpan}
\ea
\\
where, $P^{ij}_{T}=\delta^{ij}-k^{i}k^{j}/{k^{2}}$,
$P^{ij}_{L}=k^{i}k^{j}/{k^{2}}$, $P^{ij}_{n}={\tilde{n}}^i{\tilde{n}}^j/{\tilde{n}}^2$ 
and $P^{ij}_{kn}=k^{i}{\tilde{n}}^{j}+k^{j}{\tilde{n}}^{i}$
\cite{Romatschke:2003ms, Romatschke:2004jh}, where, $\tilde{n}^i=(\delta^{ij}-\frac{k^i k^j}{k^2})\hat{n}^j$
 is a vector orthogonal to, $k^i$ ,{\it i.e.,}   $\mathbf{\tilde{n}}\cdot{\mathbf{k}}=0$.
The structure functions $\alpha$, $\beta$, $\gamma$ and $\delta$, can be
determined by taking the appropriate projections of the
Eq.(\ref{seexpan}) as follows,  
\ba
\alpha&=&({P}^{ij}_{T}-{P}^{ij}_{n})\Pi^{ij}, ~~~\beta={P}^{ij}_{L}\Pi^{ij} \nonumber,\\
\gamma&=&(2{P}^{ij}_{n}-{P}^{ij}_{T})\Pi^{ij},~~~\delta=\frac{1}{2 k^{2}{\tilde{n}}^2}{P}^{ij}_{kn}\Pi^{ij}.
\label{structurefunctions}
\ea 

The structure functions mainly depend on $k$, 
$\omega$, $\xi$ $\nu$ and $\mathbf{k\cdot{\hat{n}}}=\cos\theta_n$. 
In the small anisotropy limit, $\xi<1$, using Eq.\eqref{pimunu}, all the structure functions can be calculated analytically up to linear order in $\xi$ as,
\begin{widetext}
\ba
\alpha \left(K\right)&=&\frac{m_D^2}{48 k}\Bigg(24 k z^2-2 k \xi  \left(9 z^4-13 z^2+4\right)+2 i \nu  z \left(\xi  \left(9 z^2-7\right)-12\right)-
2 \xi \cos{2\theta _n}\Big(k \big(15 z^4-19 z^2+4\big)\nn
&&+i \nu z\big(13-15 z^2\big)\Big)+\left(z^2-1\right)\big(3 \xi \left(k z \left(5 z^2-3\right)+i \nu  \left(1-5 z^2\right)\right)\cos{2\theta _n} + k z \left(-7 \xi +9 \xi  z^2-12\right)\nn
&&+i \nu  \left(\xi -9 \xi  z^2+12\right)\big)\ln\frac{z+1}{z-1}
\Bigg),\label{eq:alp}
\ea
\ba
\beta\left(K\right)&=&-\frac{m_D^2}{k}~\frac{2 (k z-i \nu )^2 }{ \nu  \ln \frac{z+1}{z-1}+2 i k} ~
\Bigg(1-\frac{1}{2} z \ln\frac{z+1}{z-1}+\frac{1}{12}\xi\left(1+3 \cos{2\theta _n}\right)\left(2-6 z^2+\left(3 z^2-2\right)z\ln {\frac{z+1}{z-1}}\right)\Bigg),\nn\label{eq:beta}
\ea
\ba
\gamma \left(K \right)=-\frac{m_D^2}{12 k}{\xi\left(k \left(z^2-1\right)-i \nu  z\right)\left(4-6 z^2+3 \left(z^2-1\right) z \ln\frac{z+1}{z-1}\right) \sin ^2{\theta _n}},
\label{eq:gamma}
\ea
\ba
\delta\left(K\right)&=& \frac{m_D^2}{24 k^2 }\xi~ \frac{(k z-i \nu ) }{2 k-i \nu  \ln\frac{z+1}{z-1}}\Bigg(k \left(88 z-96 z^3\right)+8 i \nu  \left(6 z^2-1\right)+\ln\frac{z+1}{z-1}\nn
&&\times\left(12 k \left(4 z^4-5 z^2+1\right)-10 i \nu  z-3~ i \nu  \left(4 z^4-5 z^2+1\right) \ln\frac{z+1}{z-1}\right)\Bigg)\cos{\theta _n}, \label{eq:delta}
\ea
\end{widetext}
where $z=\frac{\omega+i\nu}{k}$, and\be
\ln\frac{z+1}{z-1} =  \ln\frac{|z+1|}{|z-1|}+ i \bigg[arg\bigg(\frac{z+1}{z-1}\bigg) +2\pi N \bigg].
\ee
where $N$- corresponds to the number of Riemannian sheets. Now, $[\Delta^{-1}(K)]^{ij}$ given in Eq.(\ref{invprop}) can be rewritten   as,

\ba
[\Delta^{-1}(K)]^{ij}&=&(k^{2}-{\omega}^{2}+\alpha){P}^{ij}_{T}+(-{\omega}^{2}+\beta){P}^{ij}_{L}\nn
&&+\gamma{P}^{ij}_{n}+\delta{P}^{ij}_{kn}.
\label{invpropexpan}
\ea

\begin{figure*}[ht!]  
	\includegraphics[height=5cm,width=7.8cm]{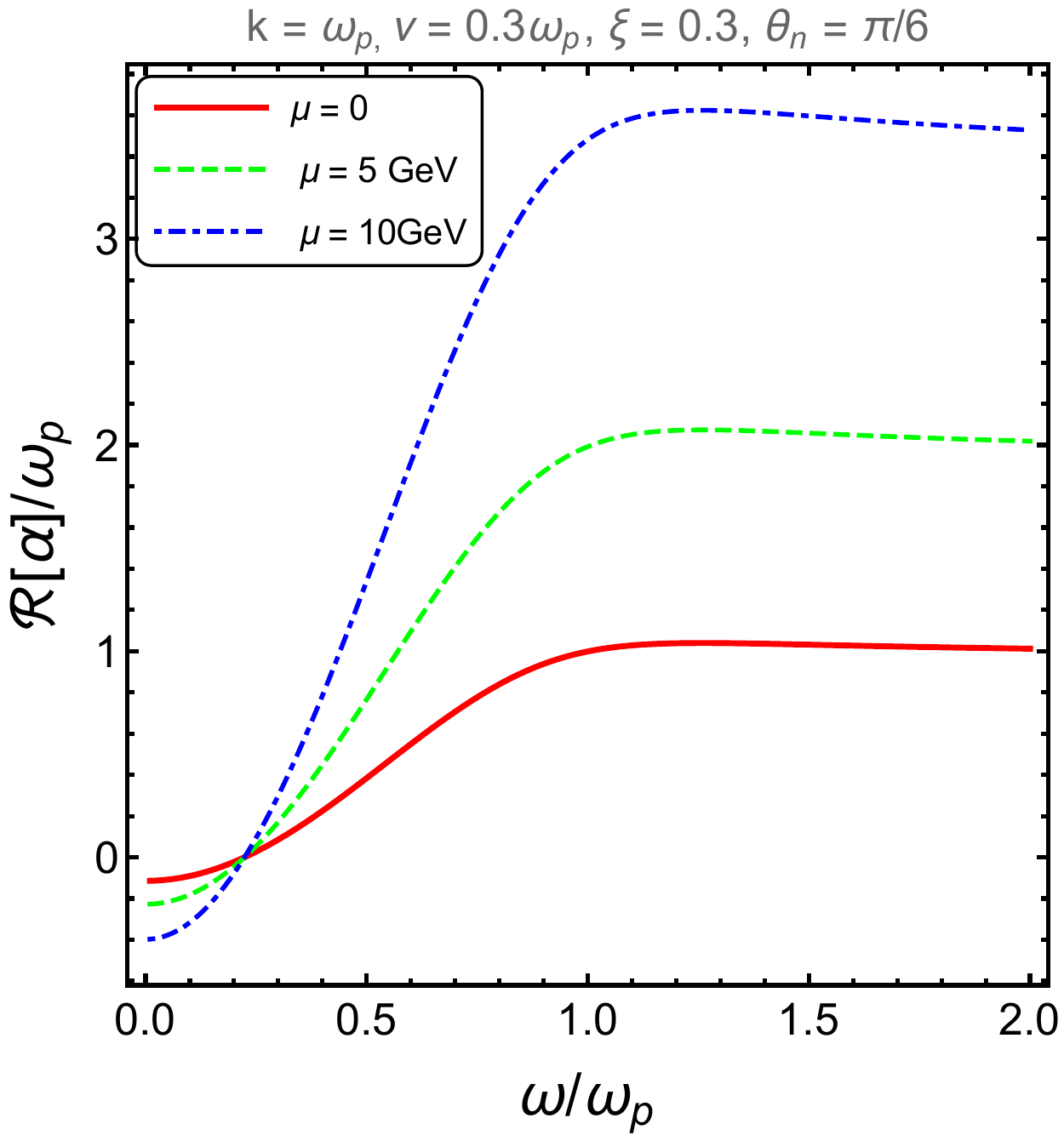}
	\hspace{3mm}
	\includegraphics[height=5cm,width=7.8cm]{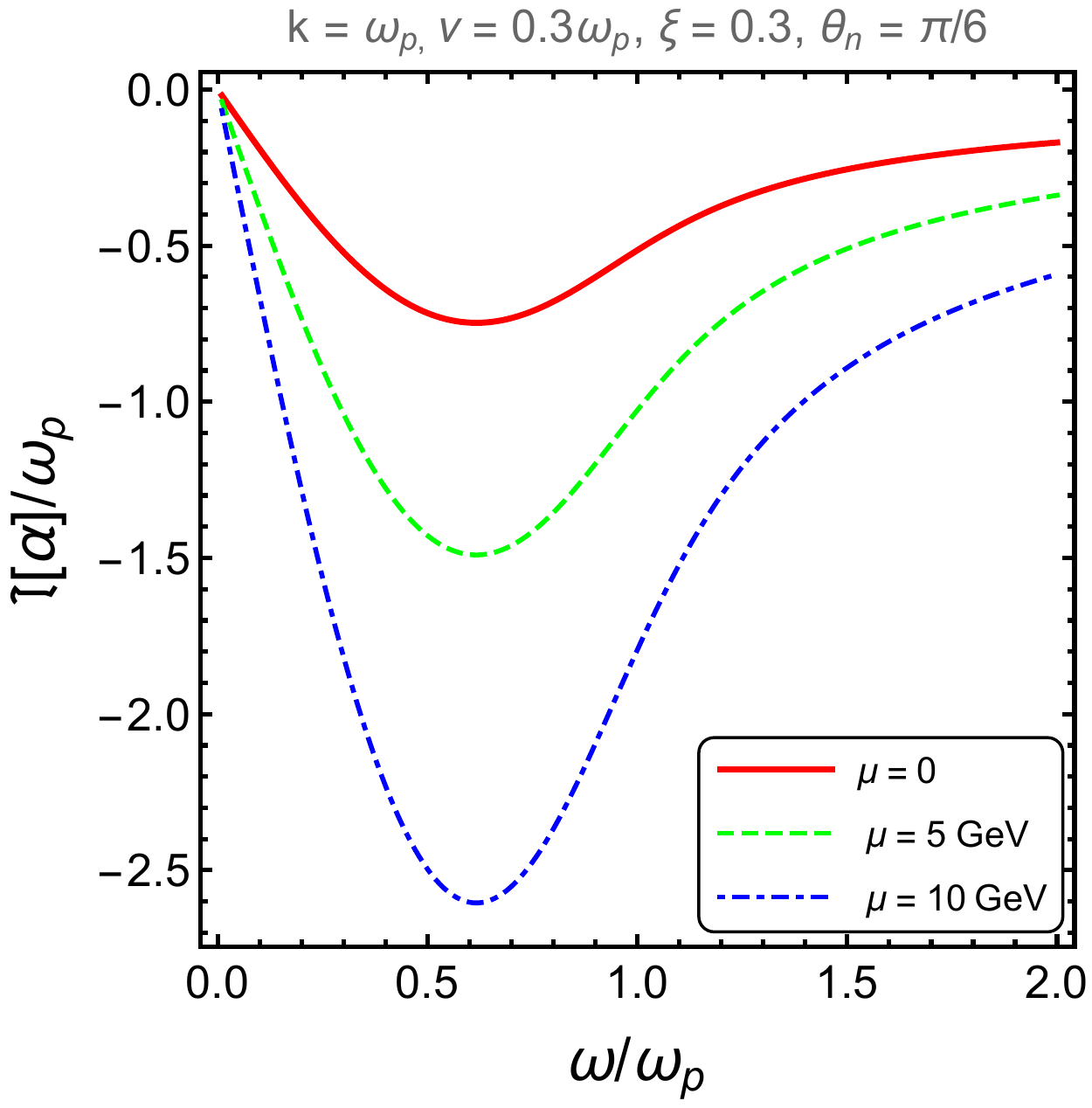}
	\vspace{3mm}
\includegraphics[height=5cm,width=7.8cm]{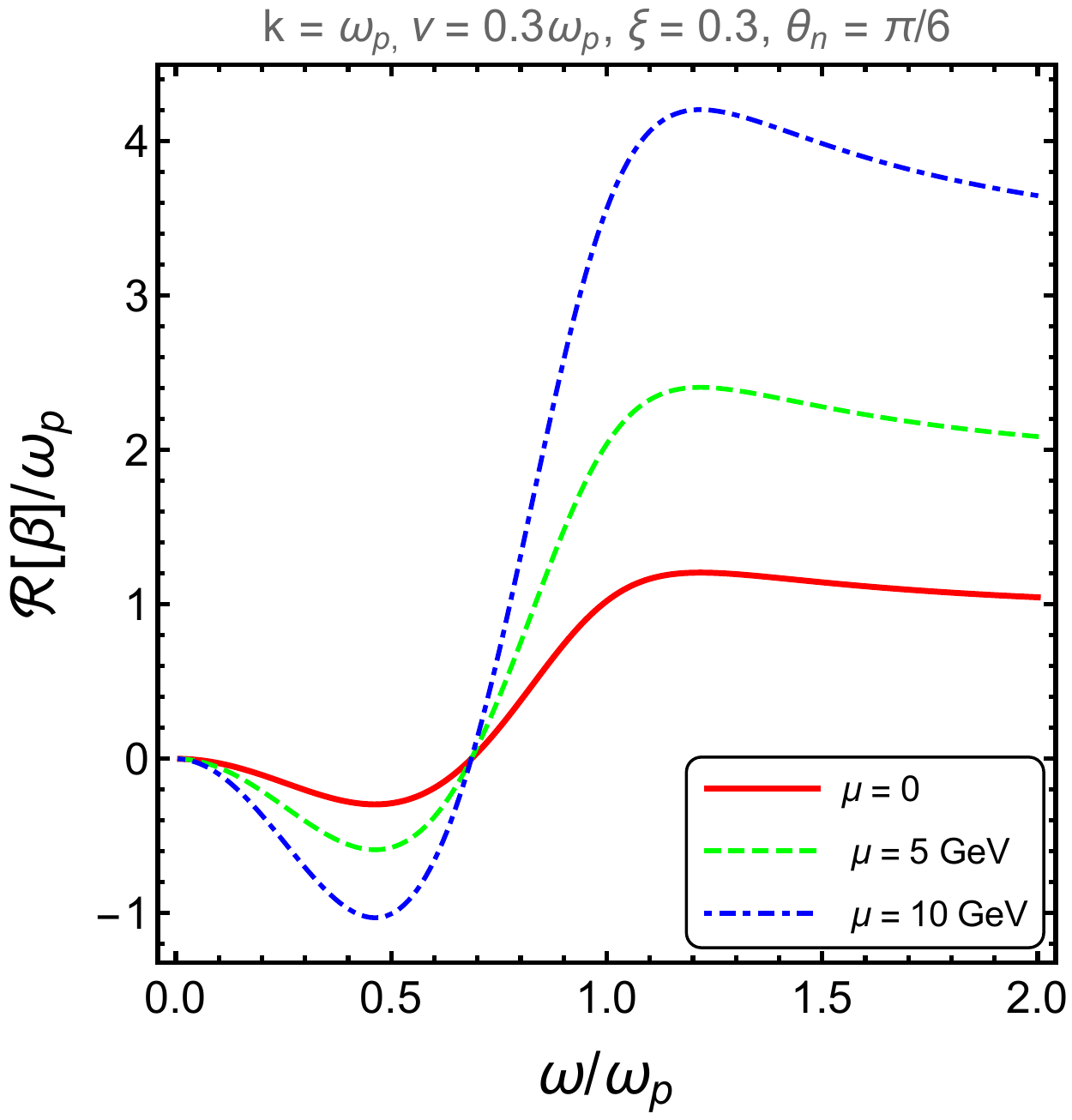}
	\hspace{3mm}
	\includegraphics[height=5cm,width=7.8cm]{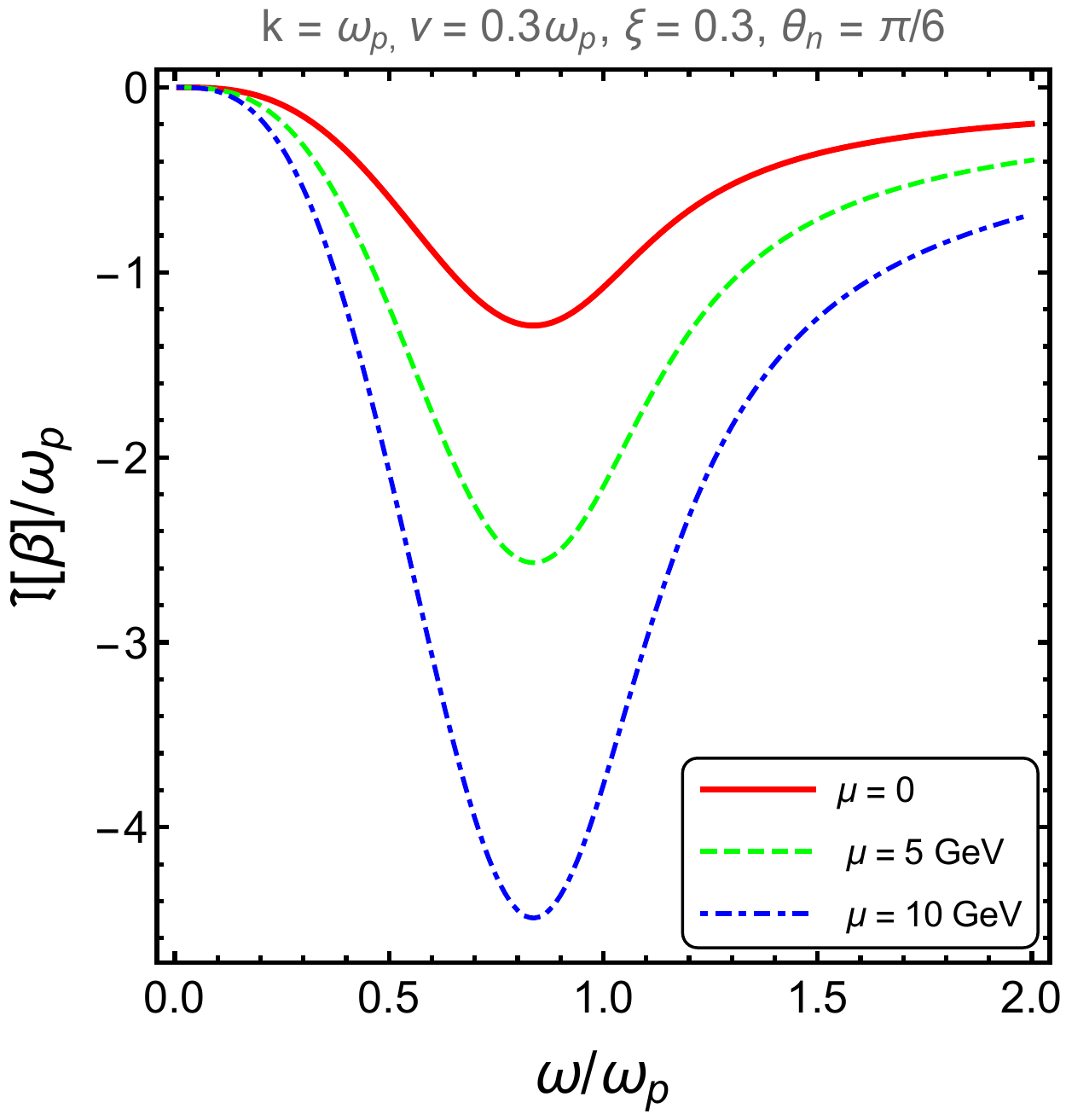}
	\vspace{3mm}
 \includegraphics[height=5cm,width=7.8cm]{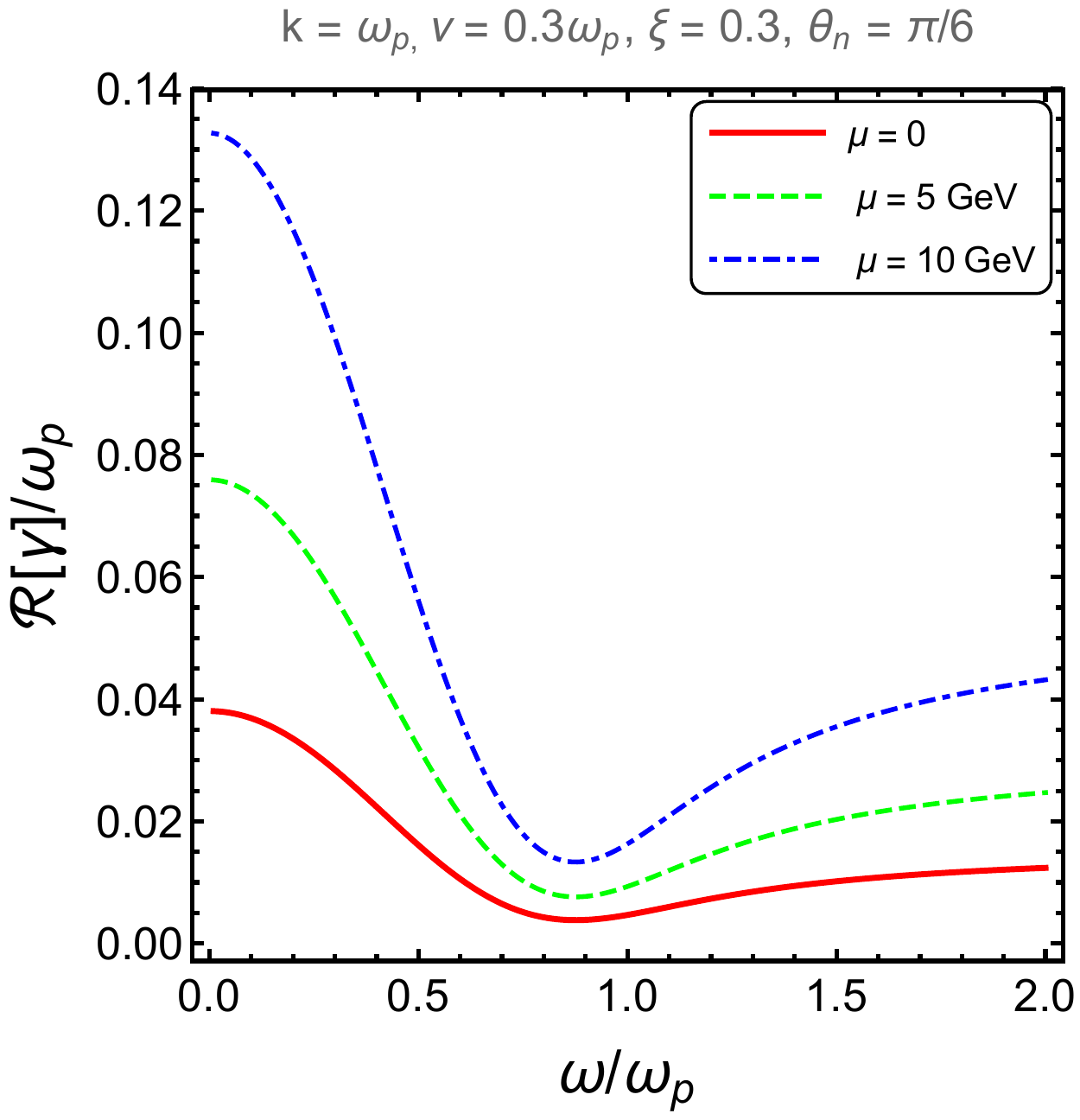}
	\hspace{3mm}
	\includegraphics[height=5cm,width=7.8cm]{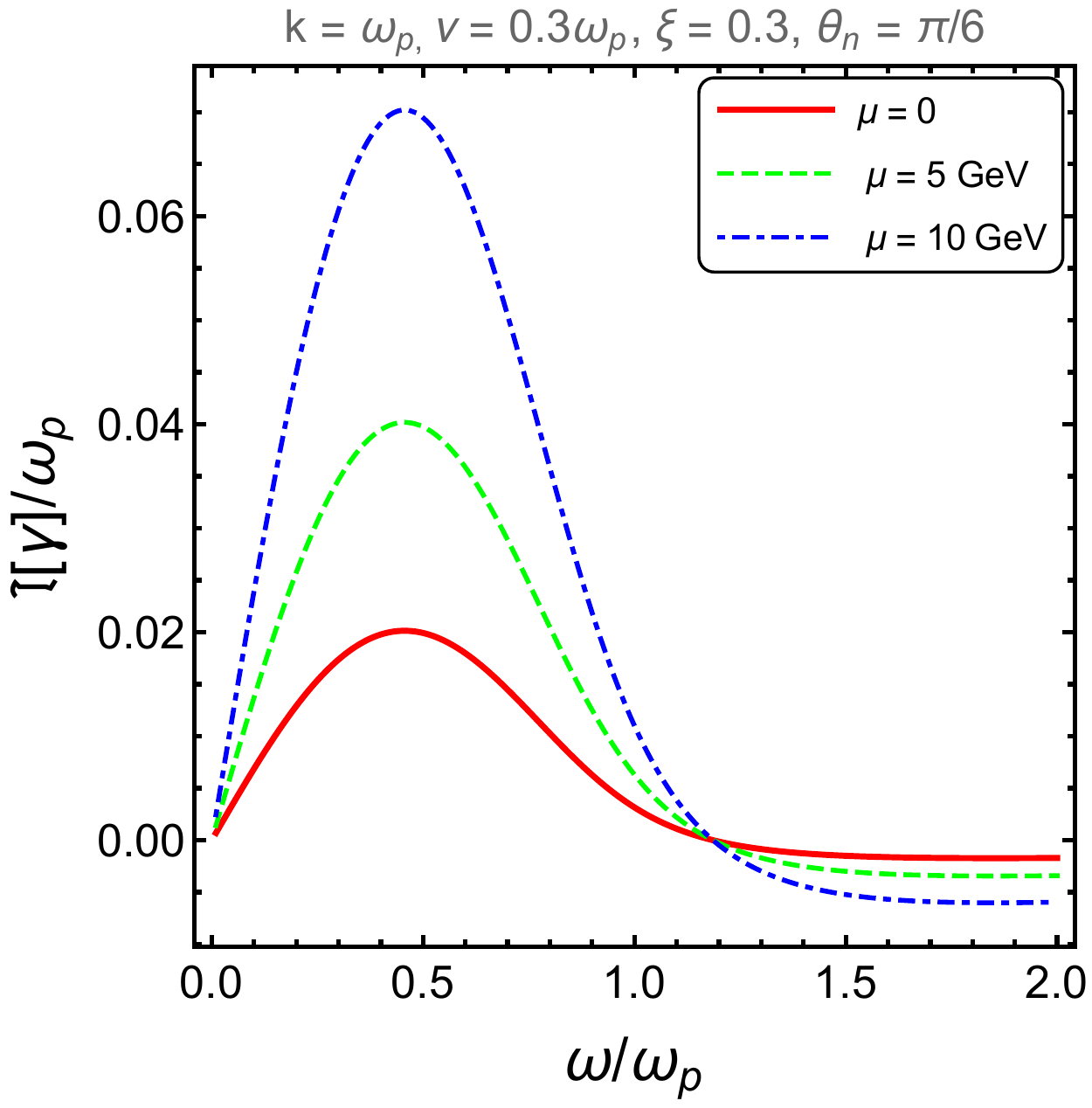}
	\vspace{3mm}
 \includegraphics[height=5cm,width=7.8cm]{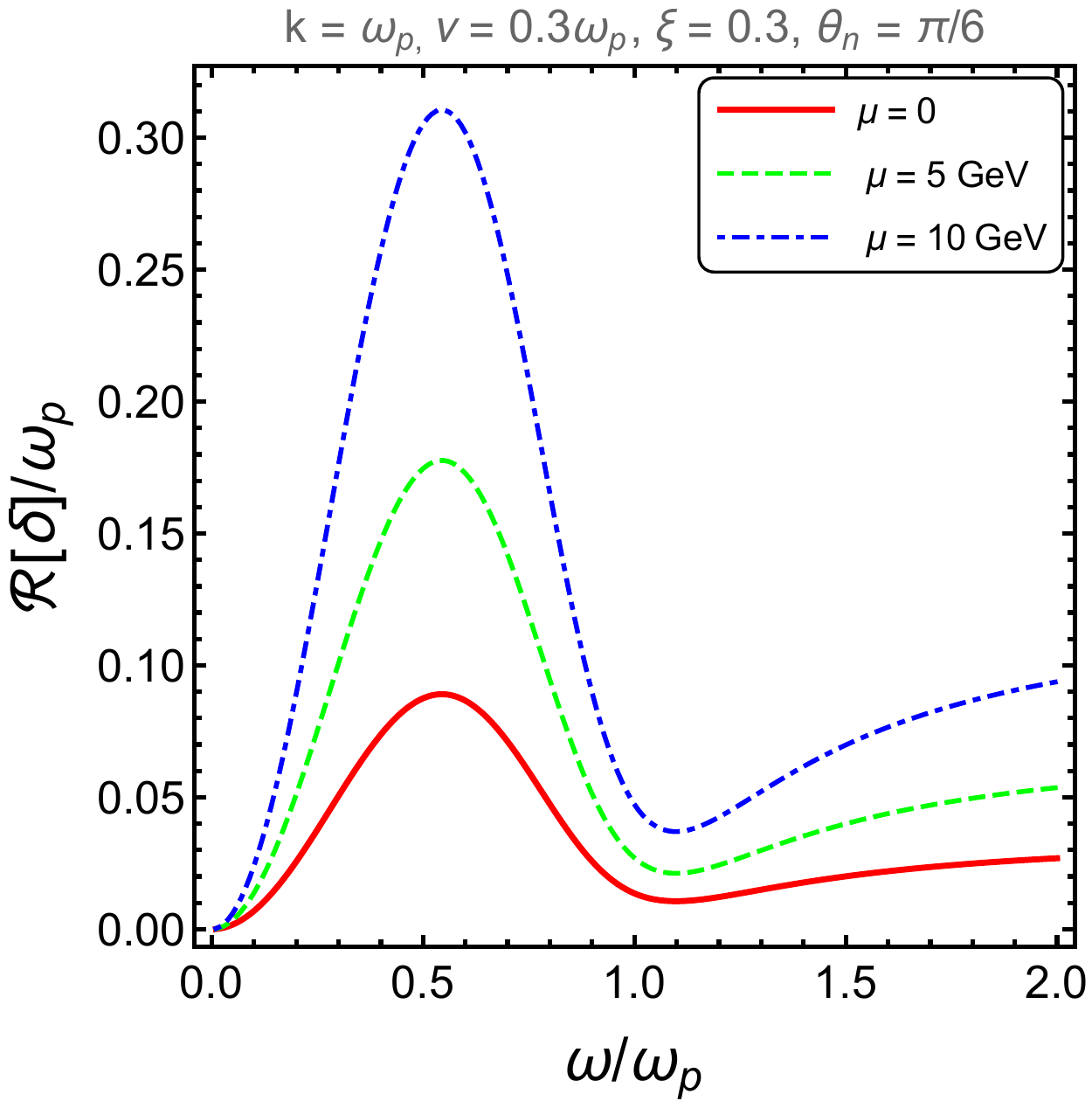}
	\hspace{3mm}
	\includegraphics[height=5cm,width=7.8cm]{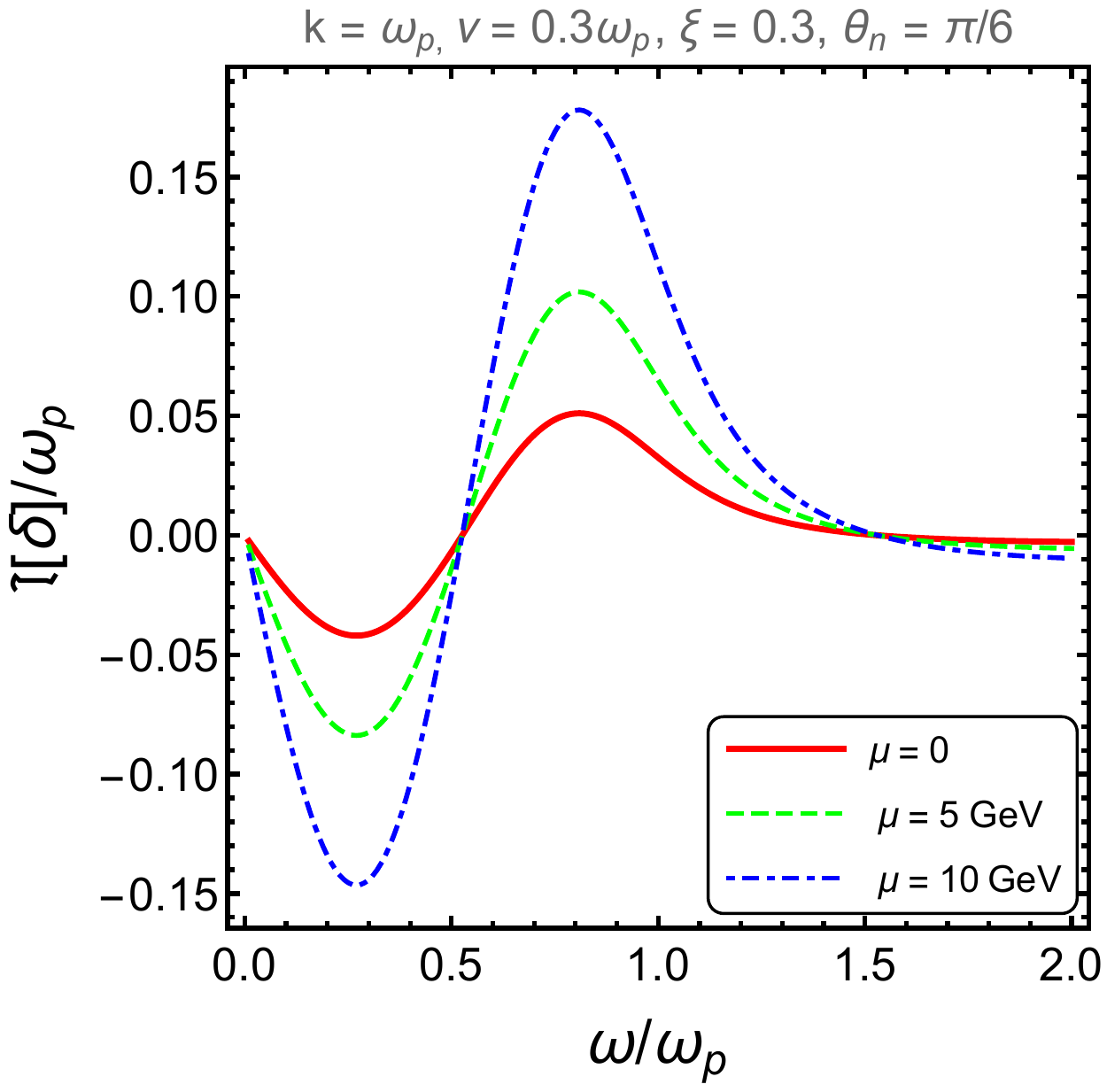}
 \caption{{Variation of real parts (left panel) and imaginary parts (right panel)  of structure functions at $\nu =0.3~\omega_p$, $\xi = 0.3$, $\theta_n=\pi/6$ and $T = 1.5~T_c$ where $T_{c} = 0.155$ GeV at different $\mu$.}}
 \label{fig:str}
\end{figure*}

\begin{figure*}[ht!]      
	\includegraphics[height=5cm,width=5.6cm]{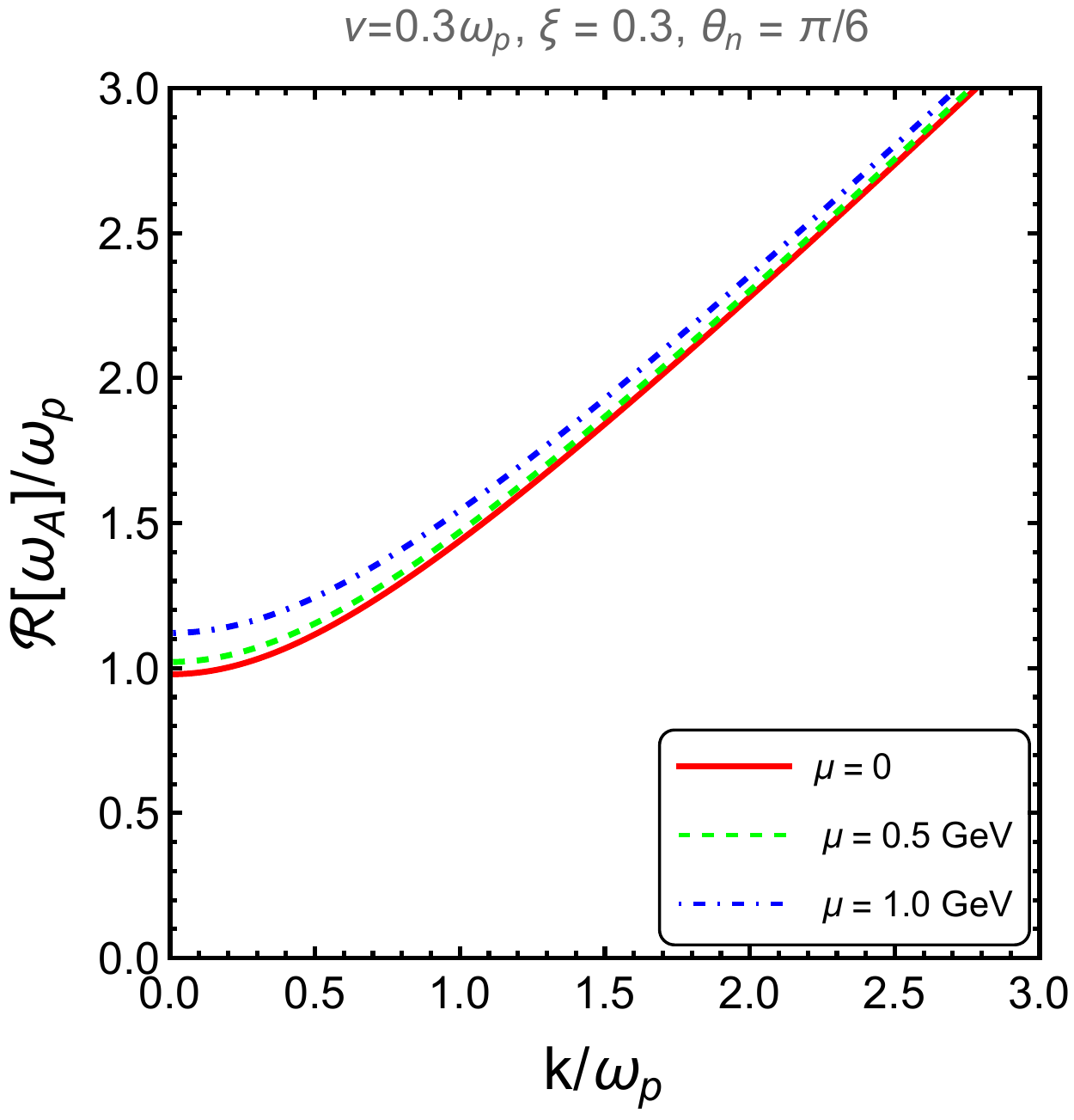}
	\hspace{3mm}
	\includegraphics[height=5cm,width=5.6cm]{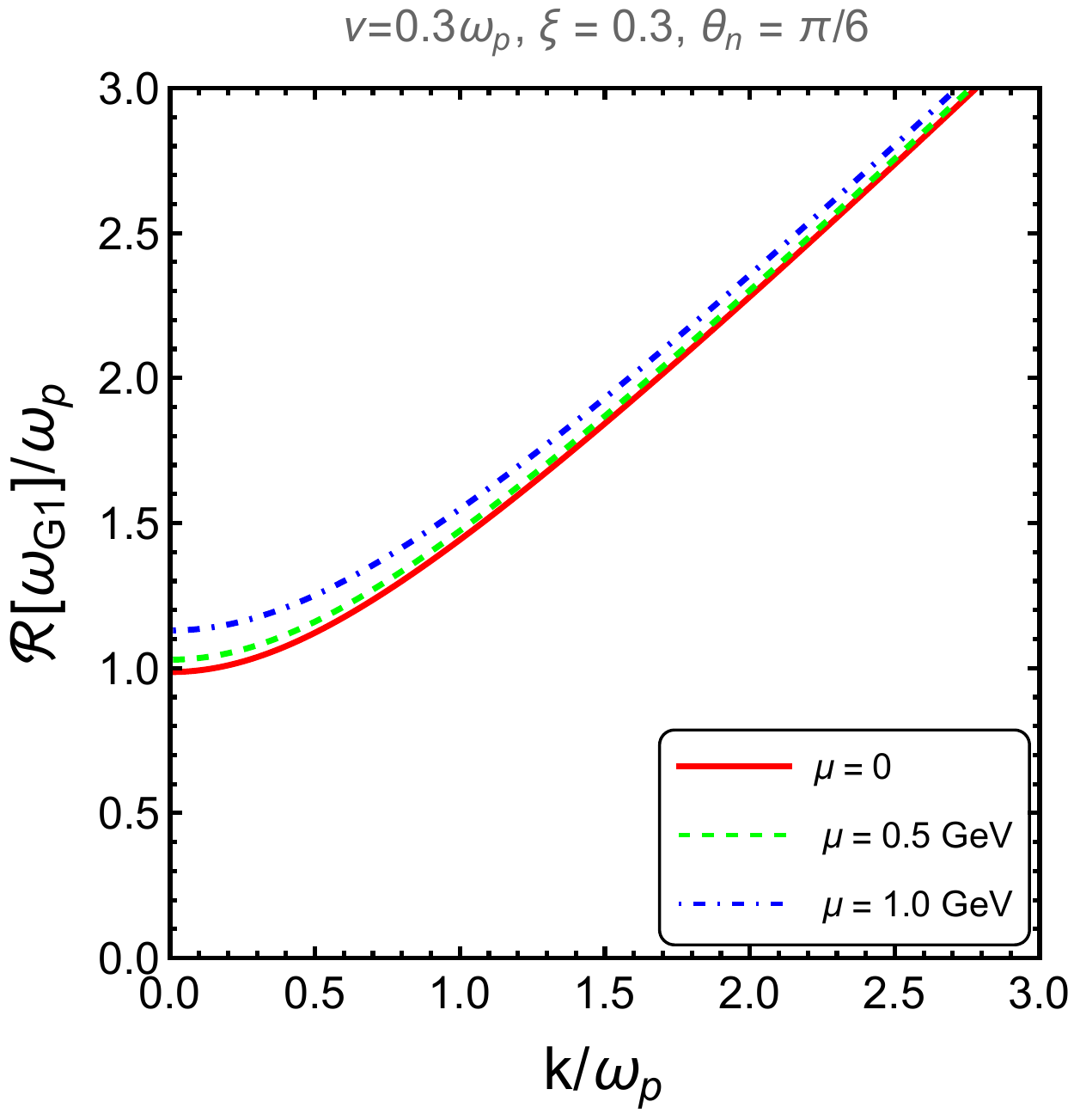}
	\vspace{3mm}
        \includegraphics[height=5cm,width=5.6cm]{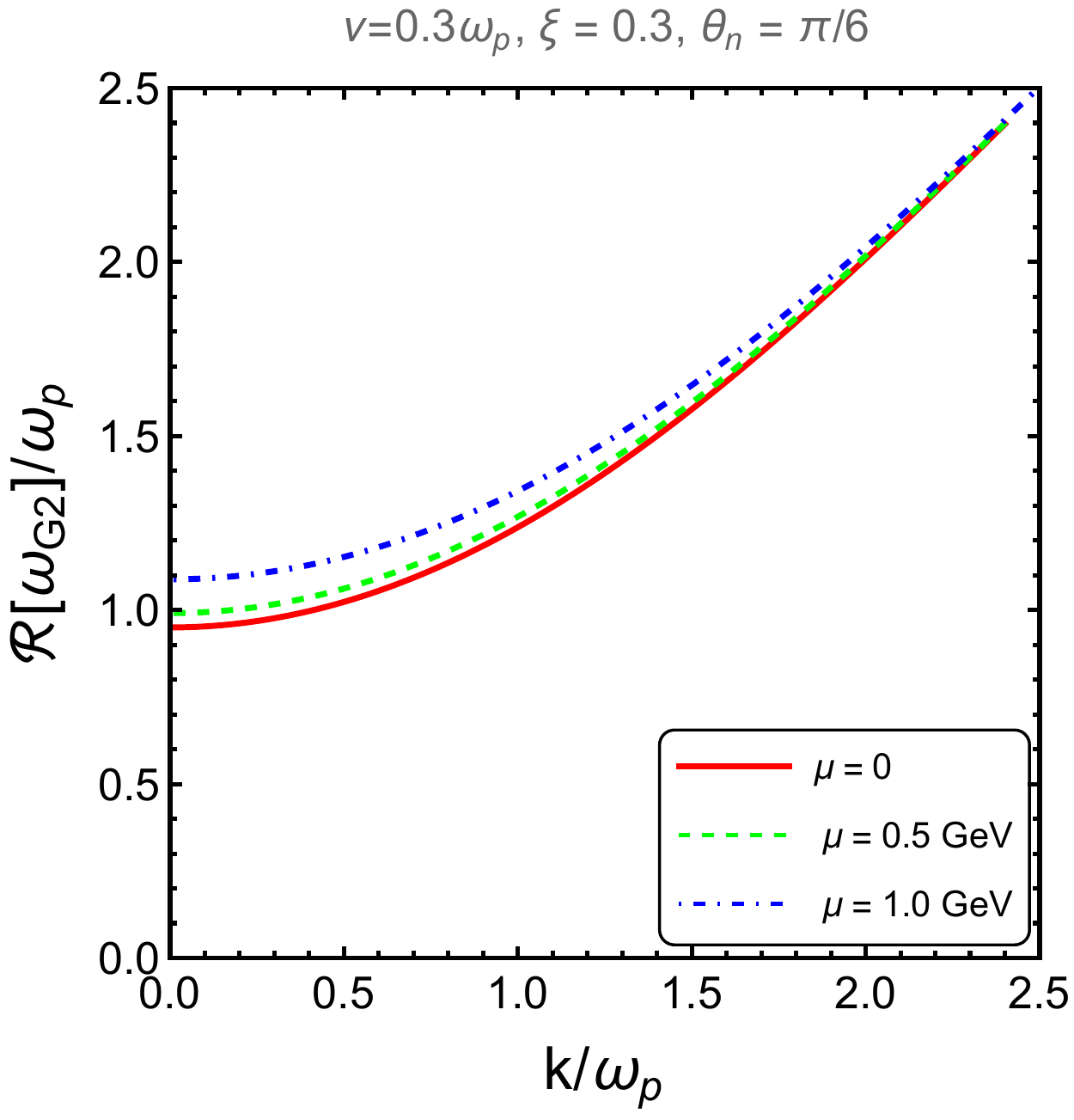}
        \caption{{Variation of real stable modes at $\nu =0.3~\omega_p$, $\xi = 0.3$, $\theta_n=\pi/6$ and $T = 1.5~T_c$ where $T_{c} = 0.155$ GeV at different $\mu$.}}
        \label{fig:re_sta}
\end{figure*}

Next, we know that both a tensor and its inverse lie in a space spanned by the same basis vectors or projection operators. Hence,
 $[\Delta(K)]^{ij}$ can also be expanded as its inverse,
\ba
[\Delta(K)]^{ij}=a{P}^{ij}_{L}+b{P}^{ij}_{T}+c{P}^{ij}_{n}+d{P}^{ij}_{kn}, \label{propagator}
\ea
but the structure constants ($a$, $b$, $c$ and $d$) would be different. To obtain that, we follow the relation 
$[\Delta^{-1}(K)]^{ij} [\Delta(K)]^{jl}=\delta^{il}$, we get,
\ba
a= \Delta_G(k^2-\omega^2+\alpha+\gamma),~~~~~~~~~~
b=\Delta_A\nn
c=\Delta_G(\beta-\omega^2)-\Delta_A,~~~~~~~~~~
d=-\Delta_G\delta
\ea
where, 
\ba
\Delta_A&=&(k^2-\omega^2+\alpha)^{-1}\nn
\Delta_G&=&[(k^2-\omega^2+\alpha + \gamma)(\beta-\omega^2)-k^2 \tilde n^2 \delta^2]^{-1}.
\ea
Considering the linear $\xi$, approximation we ignore 
$\delta^2$, as it will be of order $\xi^2$,  we have
\ba
\Delta_G=[(k^2-\omega^2+\alpha + \gamma)(\beta-\omega^2)]^{-1}.
\ea
Rewriting,  Eq.~\eqref{propagator} as,
\ba
[\Delta(K)]^{ij}&=&\Delta_A({P}^{ij}_{T}-{P}^{ij}_{n})+\Delta_G\big[(k^2-\omega^2+\alpha+\gamma){P}^{ij}_{L}\nn
&&+(\beta-\omega^2){P}^{ij}_{n}-\delta {P}^{ij}_{kn}\big]\,\text{,}
\label{propagator1}
\ea
This is a simplified structure of the gluon propagator that could be used to search the poles.

\subsection{Modes dispersion relation}
As noted earlier, the dispersion relation of the collective modes can be acquired from the poles of the gluon propagator. The poles of Eq. ~\eqref{propagator1} are given as,
\ba
\Delta^{-1}_{A}(K)=k^2-\omega^2+\alpha=0,\label{mode_a}
\ea
\ba
\Delta_G^{-1}(K) = (k^2 - \omega^2 + \alpha + \gamma)(\beta-\omega^2)=0.
\label{eq:Gx}
\ea
Eq.\eqref{eq:Gx} can further splits as,
\ba
\Delta_G^{-1}(K) = \Delta_{G1}^{-1}(k) ~\Delta_{G2}^{-1}(k)=0.
\ea
Thus, we have two more dispersion equations,
\ba
\Delta_{G1}^{-1}(K) = k^2 - \omega^2 + \alpha + \gamma=0,\label{mode_g1}\\
\Delta_{G2}^{-1}(K) = \beta-\omega^2=0.\label{mode_g2}
\ea
Note that we finally got three dispersion equations (\ref{mode_a}), (\ref{mode_g1}), and (\ref{mode_g2}). 
We call these A-, G1-, and G2-mode dispersion equations, respectively. Based on their solutions, $\omega(k)$, we can identify if the modes are real or imaginary, stable or unstable, which we shall discuss in the next section.

\section{Results and Discussions}
\label{RD}
{Our primary focus involves observing collective modes within the anisotropic QGP medium while considering finite quark chemical potential and the medium particle collisions. The introduction of medium particle collisions serves the purpose of generating imaginary modes, which are absent when $\nu$ is set to zero. Similarly, the incorporation of anisotropy in the medium results in the generation of unstable modes, with no unstable mode observed when $\xi = 0$. Emphasizing the significance of finite chemical potential, with additional considerations for anisotropy and medium particle collisions, we keep other parameters fixed unless mentioned otherwise. Specifically, we maintain a finite collision frequency, setting $\nu$ at $0.3\omega_p$, and strength of anisotropy as $\xi = 0.3$. The angular dependence is established as $\theta_n=\pi/6$, ensuring the contribution of all structure constants. For instance, at $\theta_n=0$, $\gamma$ disappears, while at $\theta_n=\pi / 2$, $\delta$ vanishes. In our investigation, various values of chemical potential within the range, $0<\mu<1$ GeV are considered. When dealing with finite baryon density in HICs, our focus is on energies below that of RHIC or LHC, where formed plasma is certainly not weakly interacting. Thus, we choose a small temperature, setting $T$ at $1.5T_c$, where $T_c=0.155$ GeV. The strong coupling at this temperature is calculated using Eq.\eqref{eq:alpha_s} at different values of $\mu$.}

The screening mass, $m_D$, significantly influences these modes. Our initial focus is on illustrating the impact of finite $\mu$ and medium temperature on $m_D$, as depicted in Fig.~\ref{fig:mD}. Notably, the observed trend reveals an increase in $m_D$ with both temperature and chemical potential. Subsequently, to delve into the behavior of collective modes, we proceed by solving the dispersion equations (\ref{mode_a}), (\ref{mode_g1}), and (\ref{mode_g2}). For a more convenient analysis, we normalize the frequency $\omega(k)$ and wave vector $k$ using the plasma frequency in the absence of a chemical potential, denoted as $\omega_p = m_D(\mu=0)/\sqrt{3}$.

\begin{figure*}[ht!]      
	\includegraphics[height=5cm,width=5.6cm]{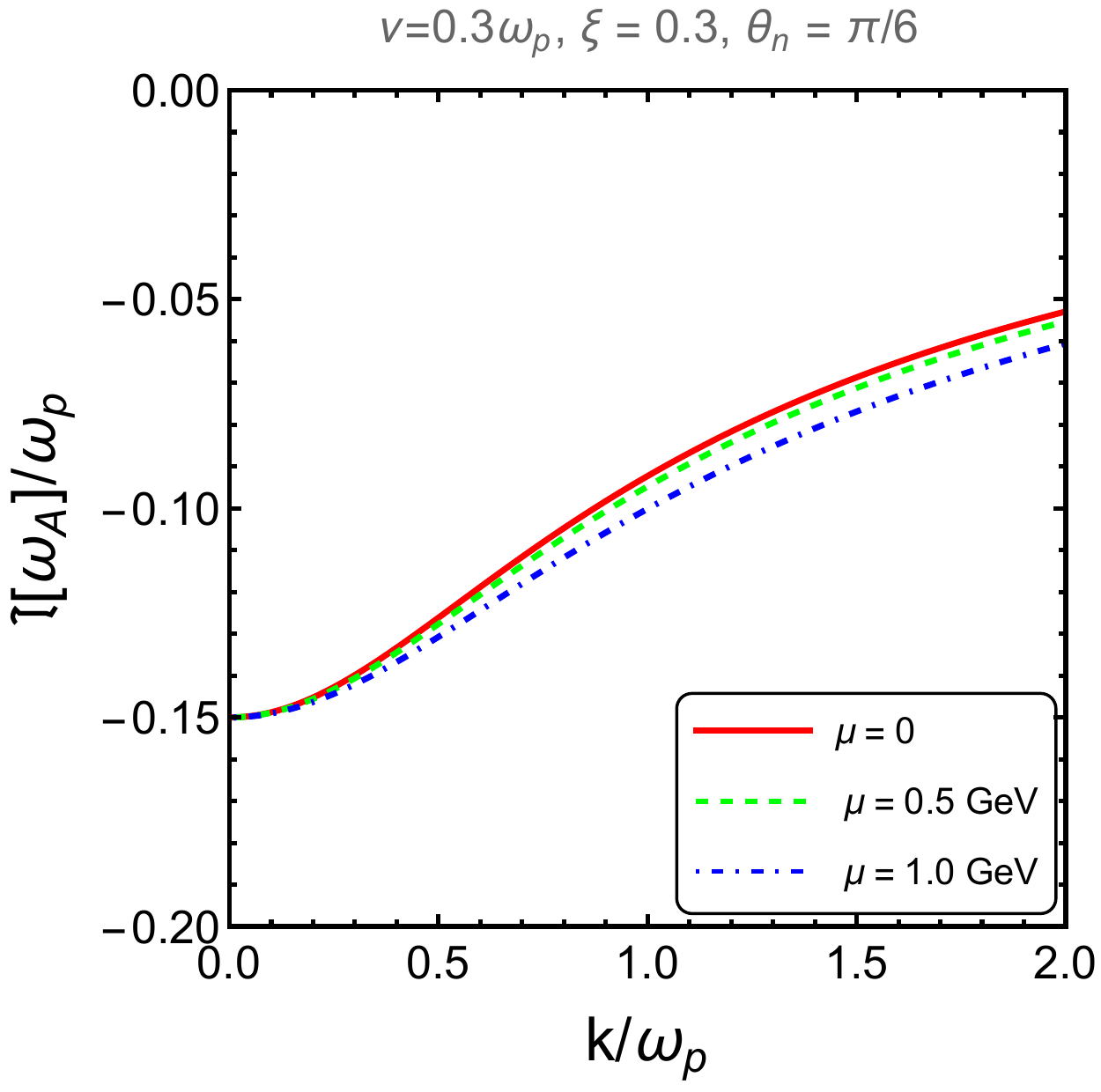}
	\hspace{3mm}
	\includegraphics[height=5cm,width=5.6cm]{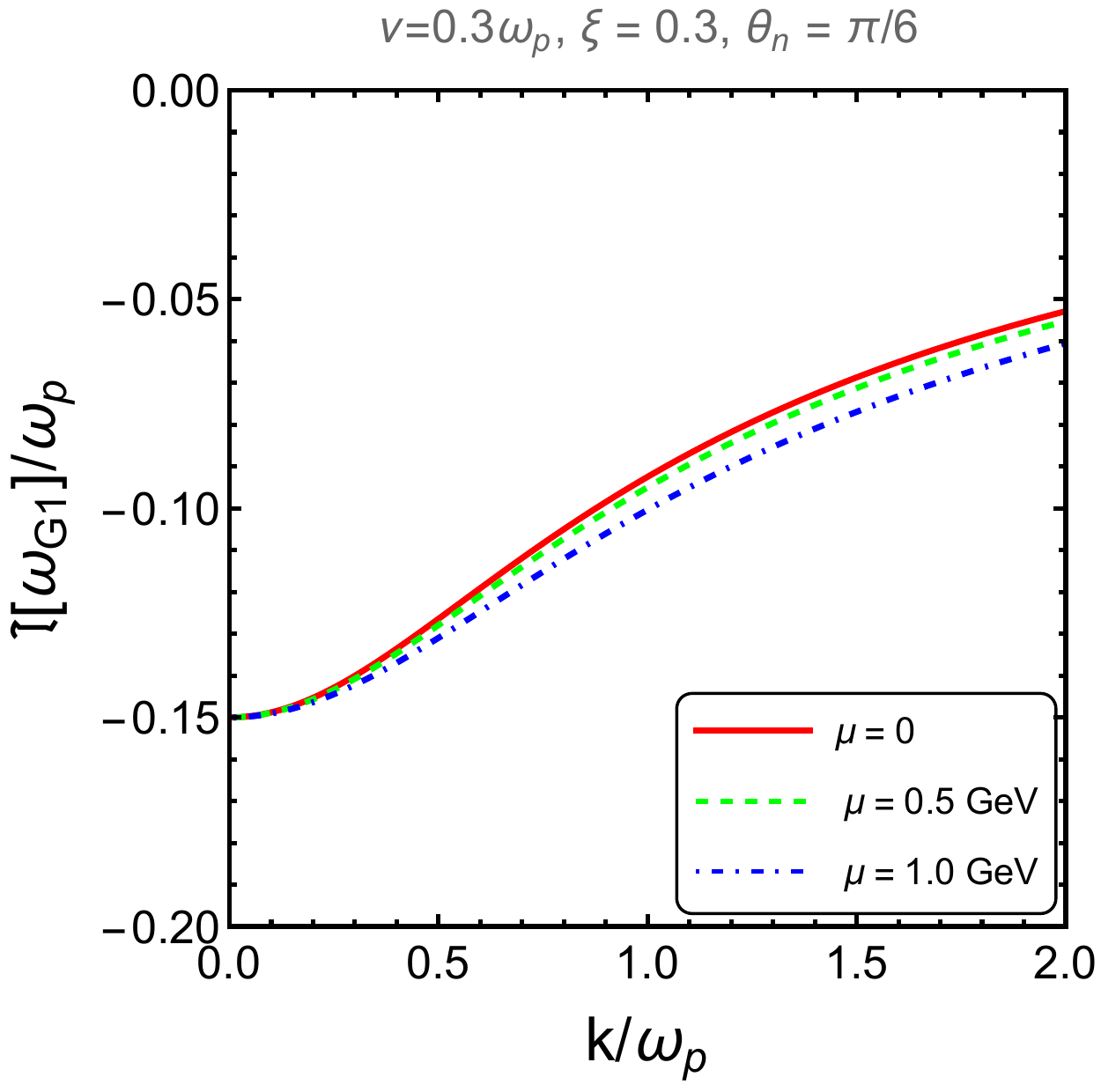}
	\vspace{3mm}
        \includegraphics[height=5cm,width=5.6cm]{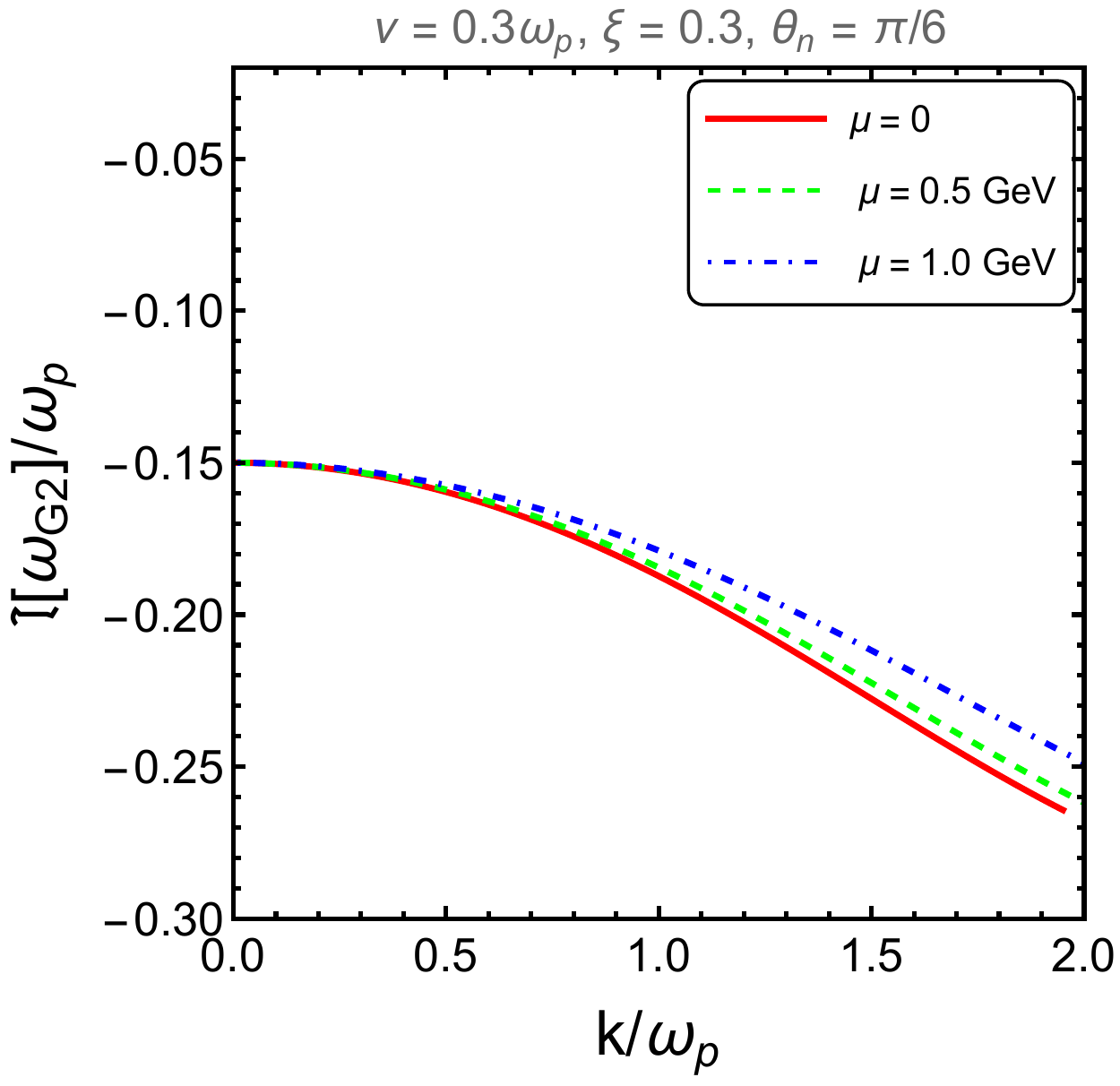}
        \caption{{Variation of imaginary stable modes at $\nu =0.3~\omega_p$, $\xi = 0.3$, $\theta_n=\pi/6$ and $T = 1.5~T_c$ where $T_{c} = 0.155$ GeV at different $\mu$.}}
        \label{fig:im_img}
\end{figure*}

\begin{figure*}[ht!]      
	\includegraphics[height=5.5cm,width=7.6cm]{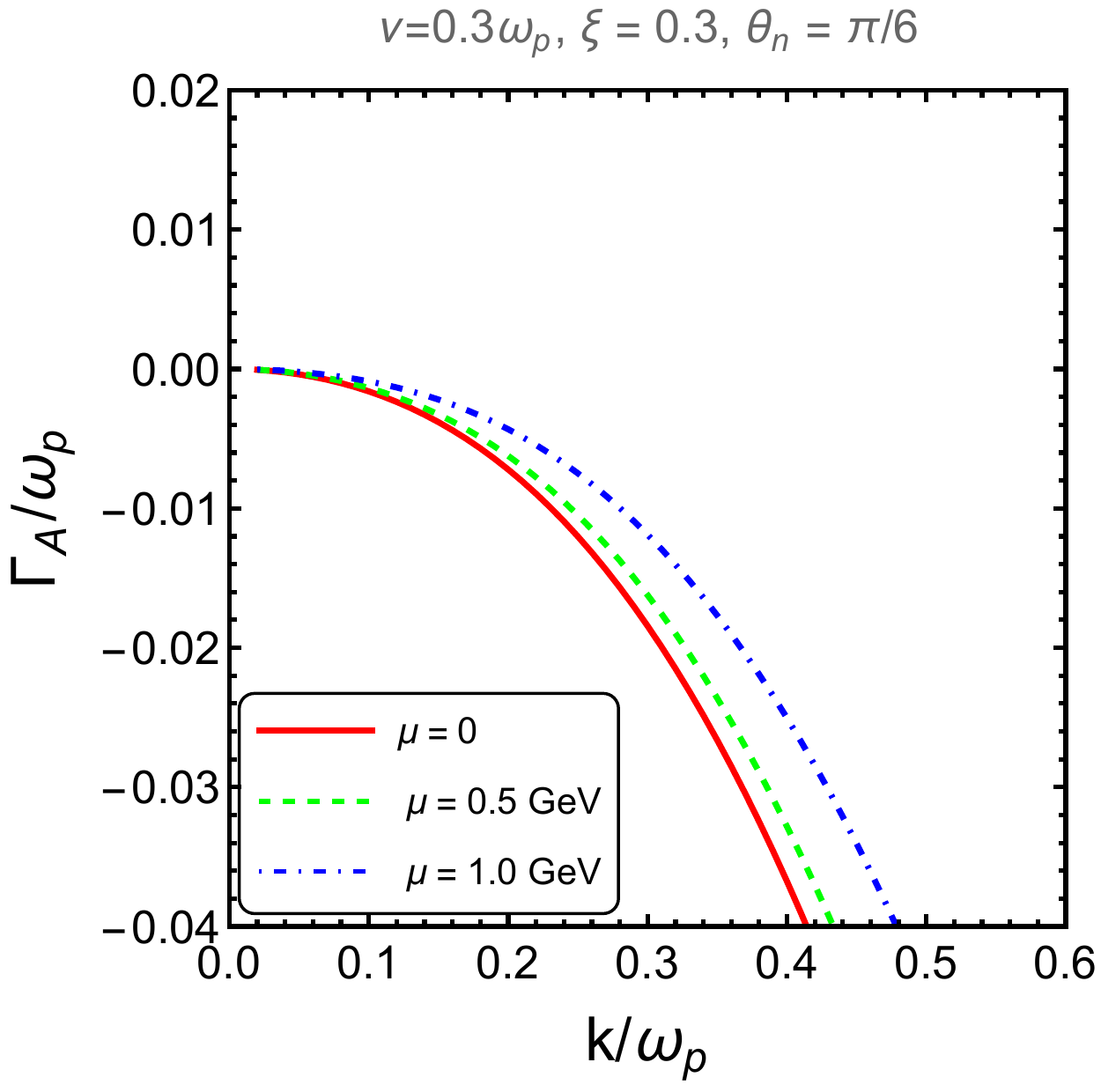}
	\hspace{3mm}
	\includegraphics[height=5.5cm,width=7.6cm]{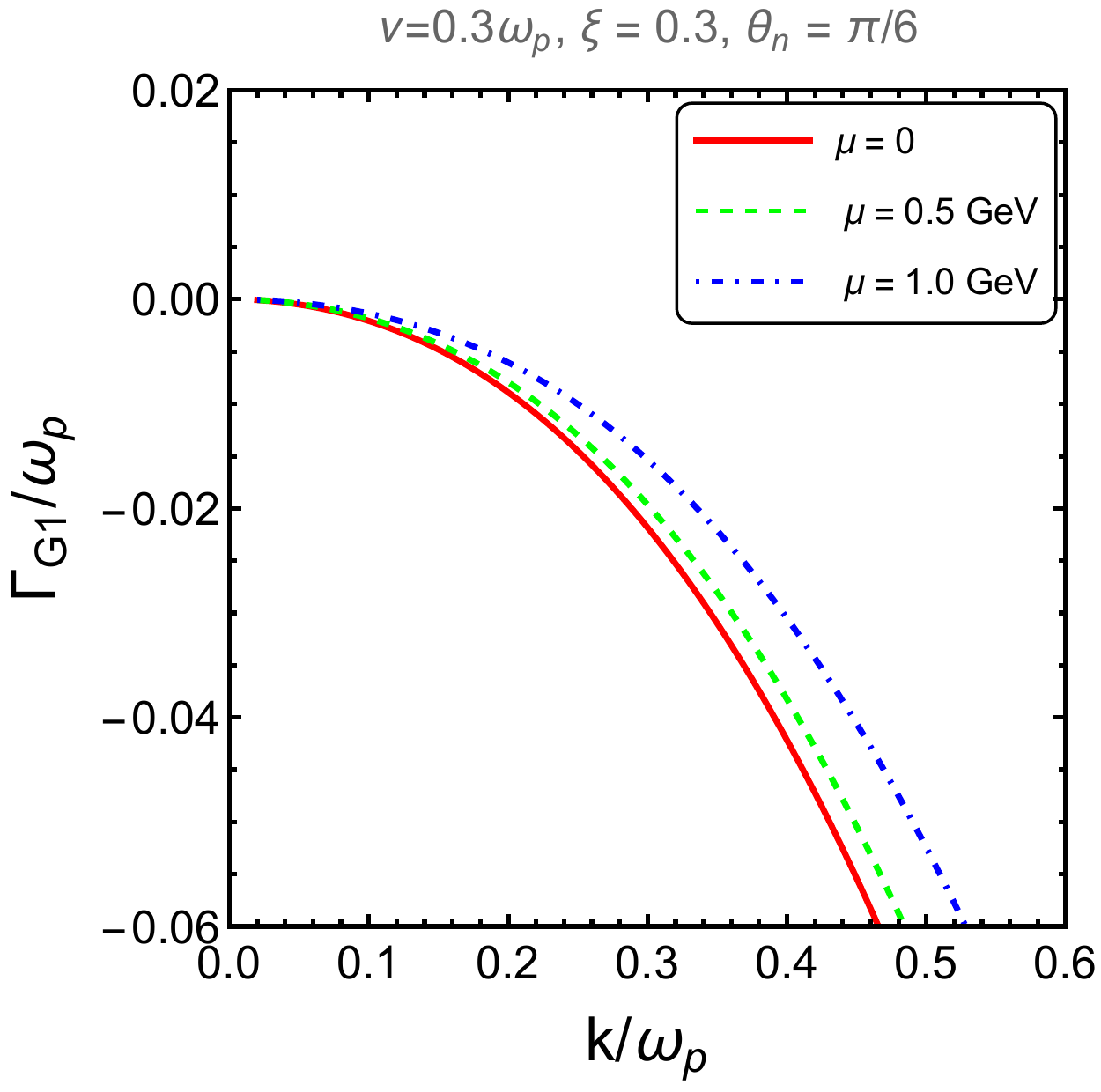}
	 \caption{{Variation of unstable modes at $\nu =0.3~\omega_p$, $\xi = 0.3$, $\theta_n=\pi/6$ and $T = 1.5~T_c$ where $T_{c} = 0.155$ GeV at different $\mu$.}}
        \label{fig:unsta}
\end{figure*}

We will begin with the discussion of the structure functions, which play a crucial role in the gluon self-energy and propagator, influencing the collective modes. The structure functions $\alpha$, $\beta$, $\gamma$, and $\delta$ given in Eq.\eqref{eq:alp} to Eq.\eqref{eq:delta} are plotted as functions of normalized frequency $\omega/\omega_{p}$ at various chemical potentials while keeping collision frequency and momentum anisotropy fixed, as shown in Fig.~\ref{fig:str}. It is evident that higher chemical potentials lead to increased magnitudes for both the real and imaginary parts of the structure functions. The behavior of these functions undergoes a change at $\omega \sim \omega_p$, where, at very high $\omega$, only the real parts of the structure functions survive, while their imaginary parts vanish. {It is to note that the magnitude of $\gamma$ and $\delta$ is almost one order less than $\alpha$ and $\beta$ at different values of $\mu$. Here, $\delta$ is no longer contributing to the dispersion relation of the modes given in Eq.\eqref{mode_a}, \eqref{mode_g1} and \eqref{mode_g2}, however, such a small value of $\gamma$ make Eq.\eqref{mode_a} and Eq.\eqref{mode_g1} almost overlapping. This is observed in various figures.}

 In Fig.~\ref{fig:re_sta}, the real stable, long-range modes are observed to increase with $k$ under the specified parameters. Notably, the stable G2-mode exhibits significant growth, while the G1-mode experiences a comparatively minute increase compared to the stable A-mode. It is worth noting that the G2-mode is considered unphysical as it violates the causality condition, $\omega \ge k$~\cite{Schenke:2006xu}.

\begin{figure*}[ht!]      
	\includegraphics[height=5.5cm,width=7.6cm]{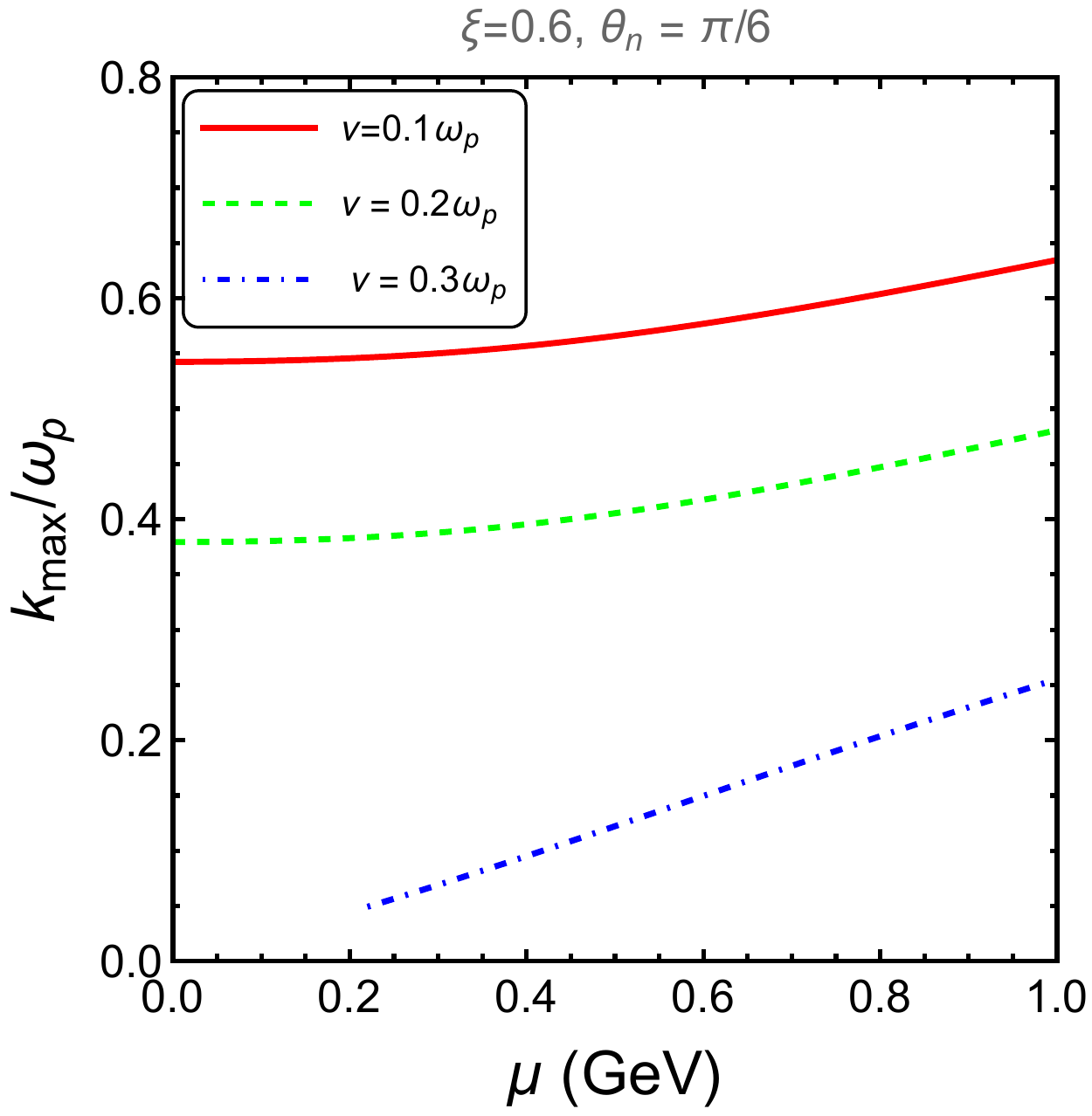}
	\hspace{3mm}
	\includegraphics[height=5.5cm,width=7.6cm]{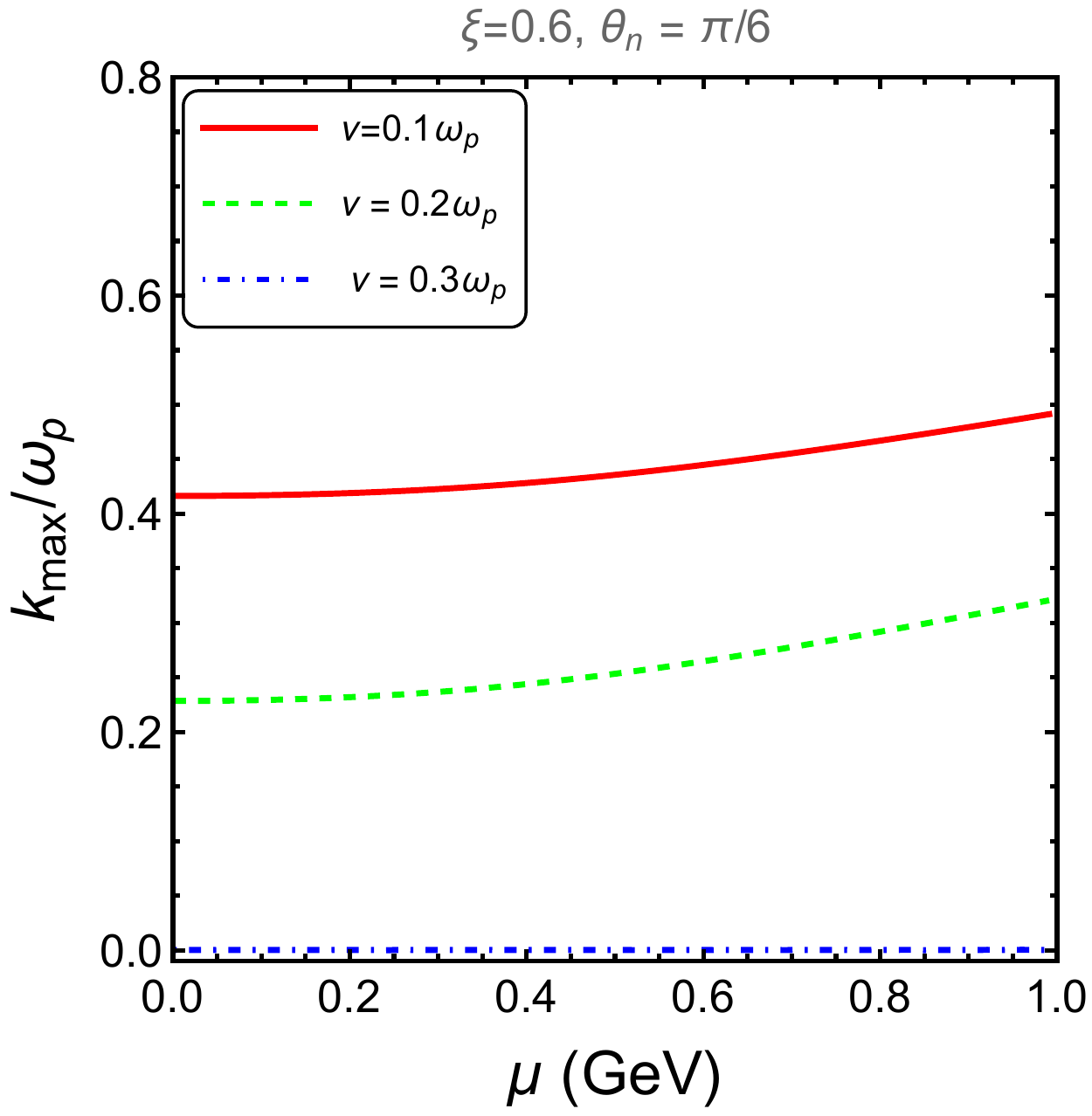}
	 \caption{{Variation of the maximum values of propagation vector for unstable A- mode (left) and G1- mode (right)  at $\xi =0.6$, $\theta_n=\pi/6$ and $T = 1.5~T_c$ where $T_{c} = 0.155$ GeV at different $\nu$.}}
        \label{fig:kmax}
\end{figure*}

The emergence of imaginary parts in the modes is primarily due to collisional effects. The finite collision frequency ($\nu$) leads to the generation of damped modes with $\Im(\omega(k))<0$, implying that these modes decay exponentially with time as $e^{-\Im(\omega(k))t}$ but do not reach the over-damped condition. In a collisionless plasma ($\nu=0$), these modes do not exist, as shown in the previous studies ~\cite{Romatschke:2003ms, Schenke:2006xu}.
Fig.~\ref{fig:im_img} illustrates the imaginary stable modes, where both A- and G1-modes exhibit an increase in magnitude with rising chemical potential. The distinction between these modes is minimal, primarily because the magnitude of $\gamma$ (which differentiates them) is relatively small compared to $\alpha$. Conversely, the imaginary G2-mode shows an opposite trend, primarily due to the absence of $k^2$ as compared to the other modes and also its dependence on $\beta$, which has a higher magnitude at a finite chemical potential.

In Fig.~\ref{fig:unsta}, two unstable modes, namely the unstable A- and G1-modes, are observed in weakly squeezed plasma, $\xi = 0.3$ and $\theta_n=\pi/6$. These are available even at $\nu=0$. However, for consistency, we considered a non-zero collision frequency, $\nu = 0.3~\omega_p$, and various chemical potential values. Here, the unstable G2-mode is absent because these modes are imaginary solutions of $\omega$ in the dispersion equations (\ref{mode_a}), (\ref{mode_g1}), and (\ref{mode_g2}), and by assuming $\omega$ to be purely imaginary ($\omega=i\Gamma$), we find that $\beta>0$ and hence, $\Gamma^2+\beta=0$ from Eq.~\eqref{mode_g2} can never be satisfied. A similar scenario is discussed in prior research in the absence of medium particle collisions~\cite{Carrington:2014bla, Jamal:2017dqs}. { These modes are short-lived, and the presence of a chemical potential enhances their magnitude, and they survive for higher values of $k$, which may affect the thermalization of the hot QCD medium.

Next, we tried to further investigate the values of anisotropic strength, medium particle collisions, and chemical potential up to which the instability (unstable modes) in a hot QCD medium survives. To do so, we sought to identify the maximum wave vector value, denoted as "$k_{max}$," at which the unstable A-mode and G1-mode are entirely suppressed. To accomplish this, we set $\omega$ to zero in the corresponding equations, as described in Eq.~\eqref{mode_a} and Eq.~\eqref{mode_g1} for the unstable A-mode and G1-mode, respectively. The results, depicted in Fig.~\ref{fig:kmax}, reveal fascinating insights. we tried for smaller values of $\xi$, but we found $k_{max}$ survive only above $\xi=0.6$ if we do not go to higher values of $\mu$ than 1 GeV. Therefore, we plotted $k_{max}$ with $\mu$ for different values $\nu$ and fixed $\xi=0.6$. Notably, $k_{max}$ exhibits a clear dependence on the chemical potential, with higher chemical potentials corresponding to greater $k_{max}$ values. However, the increase in $\nu$ suppresses  $k_{max}$ and at $\nu= 0.3 \omega_p$,  $k_{max}$ almost completely suppress.} The G1-mode, although following a similar trend, is observed to be somewhat suppressed in comparison to the A-mode, primarily due to the influence of the structure function $\gamma$. These findings elucidate the complex interplay of factors affecting instability in a hot QCD medium, offering valuable insights into the behavior of collective modes under extreme conditions.
\section{Summary and Conclusions}
\label{SC}
In this comprehensive study, our investigation encompassed the intricate interplay of the chemical potential, medium particle collisions, and momentum anisotropy, all factors that influence the behavior of collective excitations within the hot QGP medium. We first studied the derivation of the gluon self-energy, considering the finite chemical potential at the particle distribution level. The introduction of momentum anisotropy was handled adeptly, allowing for the extension or compression of the distribution function along a specific direction, denoted as ${\bf \hat{n}}$. Subsequently, the linearized Boltzmann equation, suitably modified to account for medium particle collisions through the BGK collisional kernel, was diligently solved. This yielded changes in the distribution functions for the various species, including quarks, anti-quarks, and gluons. Next, we formulated the gluon propagator using the Yang-Mills equation, akin to classical Maxwell's equations. From the poles of this propagator, we derived the dispersion equations governing the collective modes. These modes were rigorously categorized as real or imaginary and stable or unstable, all depending on the solutions to these dispersion equations, as meticulously explained in our introduction section.

In our analysis, we have flexibility in examining modes across varying angular directions, simply achieved by adjusting the parameter $\theta_n$, which represents the angle between the wave vector ${\bf k}$ and the anisotropy direction ${\bf \hat {n}}$. The presence of anisotropy was necessary in the emergence of unstable modes, a phenomenon vital for our investigation. Therefore, to simplify our presentation and focus on the key aspects, we refrained from showcasing the variation of $\theta_n$ and $\xi$. We first investigated the structure functions of the self-energy and found that they intensely depend on the chemical potential. From the poles of the propagator, we got three dispersion equations for collective modes. Based on their solutions, we found three real and imaginary stable modes. Whereas only two unstable modes were found. The imaginary stable modes disappear in the absence of medium particle collision. It has been found that the chemical potential significantly enhances the magnitude of these modes. We also investigate the suppression of unstable modes through $k_{max}$ by plotting them against $\mu$ at different values of $\nu$ and fixed $\xi$. It has been observed that with $\mu$, the maximum values of the wave vector increase, which is further enhanced with anisotropy, whereas the higher values of $\nu$ suppress the $k_{max}$.

This project can be expanded by incorporating a non-local BGK kernel, which would add complexity and accuracy to the analysis. Additionally, to make the study more realistic and closely aligned with experimental observations, we can include the non-abelian term of the field strength tensor. Furthermore, it is crucial to investigate the group velocity of these modes, as they behave like plasma waves. Therefore, a valuable extension of the current analysis would involve studying the group velocity and its implications.

\section{Acknowledgements}
 MY Jamal would like to acknowledge the SERB-NPDF (National postdoctoral fellow) File No. PDF/2022/001551. MY Jamal also thanks Dr. Avdhesh Kumar for useful discussions.\\
\\
 \\
{\bf Data Availability Statement:} No Data associated in the manuscript.

{}

\begin{thebibliography}{99}

\bibitem{expt_rhic}
J. Adams {\it et al.}  (STAR Collaboration), Nucl.  Phys.  A {\bf 757}, 102 (2005);
K. Adcox {\it et al.} PHENIX Collaboration, Nucl. Phys.  A {\bf 757}, 184 (2005);
B.B. Back {\it et al.} PHOBOS Collaboration, Nucl. Phys. A  {\bf 757}, 28 (2005);
I. Arsene {\it et al.} BRAHMS Collaboration, Nucl. Phys. A {\bf 757}, 1 (2005).

\bibitem{expt_lhc}
K. Aamodt {\it et al.} (The Alice Collaboration), Phys. Rev. Lett. {\bf 105}, 252302 (2010);
Phys. Rev.  Lett. {\bf 105}, 252301 (2010); Phys. Rev. Lett. {\bf 106}, 032301 (2011).

\bibitem{Ryu}
S. Ryu, J. F. Paquet, C. Shen, G. S. Denicol, B. Schenke, S. Jeon, C. Gale, Phys. Rev. Lett. {\bf 115}, 132301 (2015).

\bibitem{Denicol1}
G. Denicol, A. Monnai, B. Schenke, Phys. Rev. Lett. {\bf 116}, 212301 (2016).

\bibitem{Chu:1988wh}
M. C. Chu and T. Matsui, Phys. Rev. D {\bf 39}, 1892 (1989).
 
 \bibitem{Agotiya:2016bqr}
V.~K.~Agotiya, V.~Chandra, M.~Y.~Jamal and I.~Nilima, Phys.\ Rev.\ D {\bf 94}, no.9, 094006 (2016)

 
\bibitem{Jamal:2018mog}
M.~Y.~Jamal, I.~Nilima, V.~Chandra and V.~K.~Agotiya,

Phys. Rev. D \textbf{97}, no.9, 094033 (2018).

\bibitem{Jamal:2020rvh}
M.~Y.~Jamal,
Springer Proc. Phys. \textbf{250}, 135-139 (2020).

 
\bibitem{Landau:1984}
{L.D.Landau and E.M.Lifshitz}, {\it Electrodynamics of continuous media}, {Butterworth-Heinemann}, (1984).

\bibitem{Jamal:2020hpy}
M.~Y.~Jamal, S.~Mitra and V.~Chandra,
J. Phys. G \textbf{47}, no.3, 035107 (2020).

\bibitem{Carrington:2015xca}
M.~E.~Carrington, K.~Deja and S.~Mrowczynski,
Phys. Rev. C \textbf{92}, no.4, 044914 (2015).

\bibitem{Ghosh:2023ghi}
R.~Ghosh, M.~Y.~Jamal and M.~Kurian,
Phys. Rev. D \textbf{108}, no.5, 054035 (2023).

\bibitem{Koike:1991mf}
Y.~Koike, AIP Conf.\ Proc.\ {\bf 243}, 916 (1992) .

\bibitem{Jamal:2020fxo}
M.~Y.~Jamal, S.~K.~Das and M.~Ruggieri,
Phys. Rev. D \textbf{103}, no.5, 054030 (2021).

\bibitem{Jamal:2021btg}
M.~Y.~Jamal and B.~Mohanty,
Eur. Phys. J. C \textbf{81}, no.7, 616 (2021).

\bibitem{Jamal:2020emj}
M.~Y.~Jamal and B.~Mohanty,
Eur. Phys. J. Plus \textbf{136}, no.1, 130 (2021).

\bibitem{YousufJamal:2019pen}
M.~Yousuf Jamal and V.~Chandra,
Eur. Phys. J. C \textbf{79}, no.9, 761 (2019).

\bibitem{Mrowczynski:1993qm}
S.~Mrowczynski, Phys.\ Lett.\ B {\bf 314}, 118 (1993).

\bibitem{Mrowczynski:1994xv}
S.~Mrowczynski, Phys.\ Rev.\ C {\bf 49}, 2191 (1994).

\bibitem{Mrowczynski:1996vh}
S.~Mrowczynski, Phys.\ Lett.\ B {\bf 393}, 26 (1997).

\bibitem{Jamal:2017dqs} 
M.~Y.~Jamal, S.~Mitra and V.~Chandra, Phys.\ Rev.\ D {\bf 95}, 094022 (2017).

\bibitem{Kumar:2017bja}
A.~Kumar, M.~Y.~Jamal, V.~Chandra and J.~R.~Bhatt,
Phys. Rev. D \textbf{97}, no.3, 034007 (2018).

\bibitem{avdhesh}
Avdhesh Kumar, Jitesh. R. Bhatt, Predhiman. K. Kaw, Phys. Letts.  {\bf B 757}, 317-323 (2016).

\bibitem{Bellac:1996}
{M. Le. Bellac}, {\it Thermal Field Theory}, {Cambridge University Press, Cambridge, UK} (2000).

\bibitem{Weibel:1959zz}
E. S. Weibel, Phys.\ Rev.\ Lett.\  {\bf 2}, 83 (1959).

\bibitem{Mrowczynski:2004kv}
S. Mrowczynski, A. Rebhan and M.~Strickland, Phys.\ Rev.\ D {\bf 70}, 025004 (2004).

\bibitem{dm_rev1}
Daniel F. Litim, C. Manual, Phys. Rep. {\bf 364}, 451 (2002).

\bibitem{dm_rev2}
J.-P.Blaizot and E.Iancu, Phys.\ Rep.\ {\bf 359}, 355 (2002).


\bibitem{Carrington:2014bla}
M.~E.~Carrington, K.~Deja and S.~Mrowczynski, Phys.\ Rev.\ C {\bf 90},  no.3, 034913 (2014).


\bibitem{Schenke:2006xu}
Bjoern Schenke, Michael Strickland, Carsten Greiner, Markus H. Thoma, Phys. Rev. D {\bf 73}, 125004 (2006).


\bibitem{Arnold:2003rq}
P. B. Arnold, J. Lenaghan and G. D. Moore, JHEP {\bf 0308}, 002 (2003).

\bibitem{Mrowczynski:2005ki}
S. Mrowczynski, Acta Phys.\ Polon.\ B {\bf 37}, 427 (2006).

\bibitem{Romatschke:2003ms}
P. Romatschke and M.Strickland, Phys. Rev. D {\bf 68}, 036004 (2003).

\bibitem{Karmakar:2022one}
B.~Karmakar, R.~Ghosh and A.~Mukherjee,
Phys. Rev. D \textbf{106}, no.11, 116006 (2022).

\bibitem{Romatschke:2004jh}
P. Romatschke and M.Strickland Phys. Rev. D {\bf 70}, 116006 (2004).

\bibitem{Schenke:2006yp}
B. Schenke and M. Strickland, Phys.\ Rev.\ D {\bf 76}, 025023 (2007).

\bibitem{Doble:2017syb}
N.~Doble, L.~Gatignon, K.~H\"ubner and E.~Wilson,
Adv. Ser. Direct. High Energy Phys. \textbf{27}, 135-177 (2017).

\bibitem{Selyuzhenkov:2020djo}
I.~Selyuzhenkov,
J. Phys. Conf. Ser. \textbf{1685}, no.1, 012020 (2020).

\bibitem{Syresin:2022mjz}
E.~Syresin, O.~Brovko, A.~Butenko, A.~Galimov, E.~Gorbachev, V.~Kekelidze, H.~Khodzhibagiyan, S.~Kostromin, V.~Lebedev and I.~Meshkov, \textit{et al.}
JACoW \textbf{IPAC2022}, 1819-1821 (2022).

\bibitem{Nagamiya:2006en}
S.~Nagamiya,
Nucl. Phys. A \textbf{774}, 895-898 (2006).

\bibitem{Blaizot:2001nr}
J.~P.~Blaizot and E.~Iancu,
Phys. Rept. \textbf{359}, 355-528 (2002).


\bibitem{Elze:1989un}
H.~T.~Elze and U.~W.~Heinz,
Phys. Rept. \textbf{183}, 81-135 (1989).

\bibitem{Jiang:2016dkf}
Bing-feng Jiang, De-fu Hou and Jia-rong Li, Phys. Rev. D {\bf 94}, 074026 (2016).

\bibitem{Bhatnagar:1954} 
P. L. Bhatnagar, E. P. Gross, and M. Krook, Phys. Rev. {\bf 94}, 511 (1954).

\bibitem{Carrington:2004}
M. Carrington, T. Fugleberg, D. Pickering, and M. Thoma, Can. J. Phys. {\bf 82}, 671 (2004).

\bibitem{Mrowczynski:2000ed}
S.~Mrowczynski and M.~H.~Thoma, Phys.\ Rev.\ D {\bf 62}, 036011 (2000) .

\bibitem{Srivastava:2010xa}
P.~K.~Srivastava, S.~K.~Tiwari and C.~P.~Singh,
Phys. Rev. D \textbf{82}, 014023 (2010).

\bibitem{Braaten:1991gm}
E.~Braaten and R.~D.~Pisarski,
Phys. Rev. D \textbf{45}, no.6, R1827 (1992).

\bibitem{Bannur:2007tk}
V.~M.~Bannur,
Phys. Rev. C \textbf{78}, 045206 (2008).





\end{thebibliography}
\end{document}